\documentclass[conference,a4paper]{IEEEtran}
\IEEEoverridecommandlockouts

\newcommand\PicSpace{5mm}

\usepackage{bm}
\usepackage{tikz}
\usetikzlibrary{positioning, decorations.text, shapes.geometric, arrows, calc, fit, matrix, positioning, shapes, shadows, trees, mindmap, tikzmark, arrows.meta, angles, quotes, babel, spy, decorations.markings}
\usetikzlibrary{decorations.pathreplacing, calligraphy}
\usepackage{pgfplots}
\pgfplotsset{compat = 1.18}
\usepackage{url}
\usepackage[per-mode=repeated-symbol]{siunitx}
\usepackage{cite}
\usepackage{amsmath,amssymb,amsfonts}
\usepackage{graphicx}
\usepackage{textcomp}
\usepackage{xcolor}
\usepackage{bbm}

\usepackage[ruled,lined,noend]{algorithm2e}

\def\BibTeX{{\rm B\kern-.05em{\sc i\kern-.025em b}\kern-.08em
    T\kern-.1667em\lower.7ex\hbox{E}\kern-.125emX}}
\definecolor{OIblack}{rgb}{0,0,0}
\definecolor{OIorange}{rgb}{0.902,0.624,0}
\definecolor{OIcyan}{rgb}{0.337,0.706,0.914}
\definecolor{OIgreen}{rgb}{0,0.620,0.451}
\definecolor{OIyellow}{rgb}{0.941,0.894,0.259}
\definecolor{OIblue}{rgb}{0,0.447,0.698}
\definecolor{OIred}{rgb}{0.835,0.369,0}
\definecolor{OIpink}{rgb}{0.8,0.475,0.655}

\definecolor{kit-green100}{rgb}{0,.59,.51}
\definecolor{kit-green70}{rgb}{.3,.71,.65}
\definecolor{kit-green50}{rgb}{.50,.79,.75}
\definecolor{kit-green30}{rgb}{.69,.87,.85}
\definecolor{kit-green15}{rgb}{.85,.93,.93}
\definecolor{KITgreen}{rgb}{0,.59,.51}

\definecolor{KITpalegreen}{RGB}{130,190,60}
\colorlet{kit-maigreen100}{KITpalegreen}
\colorlet{kit-maigreen70}{KITpalegreen!70}
\colorlet{kit-maigreen50}{KITpalegreen!50}
\colorlet{kit-maigreen30}{KITpalegreen!30}
\colorlet{kit-maigreen15}{KITpalegreen!15}

\definecolor{KITblue}{rgb}{.27,.39,.66}
\definecolor{kit-blue100}{rgb}{.27,.39,.67}
\definecolor{kit-blue70}{rgb}{.49,.57,.76}
\definecolor{kit-blue50}{rgb}{.64,.69,.83}
\definecolor{kit-blue30}{rgb}{.78,.82,.9}
\definecolor{kit-blue15}{rgb}{.89,.91,.95}

\definecolor{KITyellow}{rgb}{.98,.89,0}
\definecolor{kit-yellow100}{cmyk}{0,.05,1,0}
\definecolor{kit-yellow70}{cmyk}{0,.035,.7,0}
\definecolor{kit-yellow50}{cmyk}{0,.025,.5,0}
\definecolor{kit-yellow30}{cmyk}{0,.015,.3,0}
\definecolor{kit-yellow15}{cmyk}{0,.0075,.15,0}

\definecolor{KITorange}{rgb}{.87,.60,.10}
\definecolor{kit-orange100}{cmyk}{0,.45,1,0}
\definecolor{kit-orange70}{cmyk}{0,.315,.7,0}
\definecolor{kit-orange50}{cmyk}{0,.225,.5,0}
\definecolor{kit-orange30}{cmyk}{0,.135,.3,0}
\definecolor{kit-orange15}{cmyk}{0,.0675,.15,0}

\definecolor{KITred}{rgb}{.63,.13,.13}
\definecolor{kit-red100}{cmyk}{.25,1,1,0}
\definecolor{kit-red70}{cmyk}{.175,.7,.7,0}
\definecolor{kit-red50}{cmyk}{.125,.5,.5,0}
\definecolor{kit-red30}{cmyk}{.075,.3,.3,0}
\definecolor{kit-red15}{cmyk}{.0375,.15,.15,0}

\definecolor{KITpurple}{RGB}{160,0,120}
\colorlet{kit-purple100}{KITpurple}
\colorlet{kit-purple70}{KITpurple!70}
\colorlet{kit-purple50}{KITpurple!50}
\colorlet{kit-purple30}{KITpurple!30}
\colorlet{kit-purple15}{KITpurple!15}

\definecolor{KITcyanblue}{RGB}{80,170,230}
\colorlet{kit-cyanblue100}{KITcyanblue}
\colorlet{kit-cyanblue70}{KITcyanblue!70}
\colorlet{kit-cyanblue50}{KITcyanblue!50}
\colorlet{kit-cyanblue30}{KITcyanblue!30}
\colorlet{kit-cyanblue15}{KITcyanblue!15}

\usepackage[capitalise]{cleveref}
\crefname{section}{Sec.}{Secs.}
\Crefname{section}{Sec.}{Secs.}
\crefname{subsection}{Sec.}{Secs.}
\Crefname{subsection}{Sec.}{Secs.}
\crefname{subsubsection}{Sec.}{Secs.}
\Crefname{subsubsection}{Sec.}{Secs.}
\crefname{algorithm}{Alg.}{Algs.}
\Crefname{algorithm}{Alg.}{Algs.}
\crefname{equation}{}{}
\Crefname{equation}{}{}

\usepackage[nonumberlist, acronym]{glossaries}
\newacronym{5G}{5G}{fifth generation cellular network technology}
\newacronym{ASIC}{ASIC}{appication-specific integrated circuit}
\newacronym{BG2}{BG2}{base graph 2}
\newacronym{BP}{BP}{belief propagation}
\newacronym{BI-AWGN}{BI-AWGN}{binary-input additive white Gaussian noise}
\newacronym{CN}{CN}{check node}
\newacronym{DD-BMP}{DD-BMP}{differential decoding with binary message passing}
\newacronym{EI}{EI}{expected improvement}
\newacronym{ELENA}{ELENA-SNN}{enlarge-likelihood-each-notable-amplitude spiking-neural-network}
\newacronym{FAMP}{FA-MP}{finite alphabet message passing}
\newacronym{FEC}{FEC}{forward error correction}
\newacronym{FER}{FER}{frame error rate}
\newacronym{FinFET}{FinFET}{Fin field-effect transistor}
\newacronym{FG}{FG}{finite-geometry}
\newacronym{GP}{GP}{Gaussian process}
\newacronym{GPR}{GPR}{Gaussian process regression}
\newacronym{IBM}{IBM}{information-bottleneck method}
\newacronym{LDPC}{LDPC}{Low-density parity-check}
\newacronym{LIF}{LIF}{leaky integrate-and-fire}
\newacronym{LI}{LI}{leaky integrate}
\newacronym{LUT}{LUT}{lookup table}
\newacronym{LLR}{LLR}{log-likelihood ratio}
\newacronym{min-RCQ}{min-RCQ}{min-sum RCQ}
\newacronym{MS}{MS}{min-sum}
\newacronym{MLE}{ML-ELENA-SNN}{multi-level ELENA-SNN}
\newacronym{NMS}{NMS}{normalized min-sum}
\newacronym{PCM}{PCM}{parity check matrix}
\newacronym{PR}{P\&R}{place \& route}
\newacronym{PVT}{PVT}{process, voltage, and temperature}
\newacronym{RCQ}{RCQ}{reconstruction, computation, and quantization}
\newacronym{SNN}{SNN}{spiking neural network}
\newacronym{SNR}{SNR}{signal-to-noise ratio }
\newacronym{SPA}{SPA}{sum-product algorithm}
\newacronym{SR}{SR}{successive relaxation}
\newacronym{VN}{VN}{variable node}
\newacronym{WLAN}{WLAN}{wireless LAN}

\begin{document}

\title{
    Optimization of Finite Alphabet Message Passing Decoding on Fully Node-Parallel Architectures\\for 5G LDPC Codes
     \thanks{{This work has received funding from the German Federal Ministry of Research, Technology and Space (BMFTR) within the project Open6GHub+ (grant agreements 16KIS2405 and 16KIS2406).}
     }
}

\author{
    \IEEEauthorblockN{
        Alexander von Bank\IEEEauthorrefmark{1}, 
        Oliver Griebel\IEEEauthorrefmark{2}\IEEEauthorrefmark{3}, 
        Eike-Manuel Edelmann\IEEEauthorrefmark{1}, 
        Norbert Wehn\IEEEauthorrefmark{2}, 
        and Laurent Schmalen\IEEEauthorrefmark{1}
    }
    \IEEEauthorblockA{
        \IEEEauthorrefmark{1}\textit{Communications Engineering Lab, Karlsruhe Institute of Technology}, 76187 Karlsruhe, Germany \\
        \IEEEauthorrefmark{2}\textit{University of Kaiserslautern-Landau (RPTU)}, 67663 Kaiserslautern, Germany\\
        \IEEEauthorrefmark{3}\textit{Creonic GmbH}, 67655 Kaiserslautern, Germany\\
        (email: \texttt{alexander.bank@kit.edu},  \texttt{oliver.griebel@creonic.com})
    }
}

\IEEEaftertitletext{\vspace{-5mm}}

\maketitle

\begin{abstract}
In this paper, we propose a novel design method for the lookup tables (LUTs) of the minimum reconstruction, computation, and quantization (min-RCQ) LDPC code decoding algorithm. 
The proposed design method uses Bayesian optimization and samples from a subspace of the design space. 
The novel design method does not require knowledge about the channel or the modulation format and uses the frame error rate (FER) as loss function. 
The design method is applicable to irregular low-density parity-check (LDPC) codes with puncturing and shortening. 
Furthermore, the design method creates a set of iteration-agnostic LUTs that can be used over all iterations, which makes this approach suitable for fully node-parallel decoder architectures.
We show that our proposed approach achieves a 38.6\,\% better energy efficiency than a state-of-the-art \gls{NMS} decoder without loss in error correction performance.
\end{abstract}

\begin{IEEEkeywords}
LDPC codes, Finite Alphabet Message Passing
\vspace*{-2mm}
\end{IEEEkeywords}

\glsresetall
\section{Introduction}

\gls{LDPC} codes~\cite{G62} are widely used in modern communication standards, including \gls{5G}~\cite{3GPP26,RK18} and \gls{WLAN}~\cite{IEEE21}, due to their excellent error-correction performance under iterative message-passing decoding. 
While the \gls{SPA} can operate close to channel capacity~\cite{RS01}, its computational complexity requires simplified algorithms such as the \gls{MS}~\cite{FMI99} and the \gls{NMS}~\cite{CDEFH2005}.
Despite their reduced computational complexity, decoder implementations remain challenging because the routing networks between \glspl{CN} and \glspl{VN} must transport high-precision messages across the pseudo-random Tanner graph~\cite{GJKHW2024}. 
\Gls{FAMP} was proposed to reduce the routing complexity while preserving the error correction performance. 
\gls{FAMP} reduces the quantization-width of the exchanged messages and replaces the node update functions, i.e., in \glspl{VN} and \glspl{CN}, by \glspl{LUT}~\cite{GBMMMTB2018}.
However, conventional \gls{FAMP} suffers from exponentially growing multidimensional \gls{LUT} sizes~\cite{TG20}. 
To circumvent large \glspl{LUT}, the \gls{RCQ} approach maps low-precision messages to a higher-precision domain for computation and back afterward~\cite{WWSB2020}.
The \gls{min-RCQ} is a flavor of \gls{RCQ} that uses the \gls{MS} \gls{CN} update rule.
The \gls{min-RCQ} achieves \glspl{FER} within $0.1$\,\unit{\dB} of floating-point \gls{BP}, without exponentially growing \gls{LUT} sizes~\cite{WWSB2020}.
Designing the reconstruction and quantization \glspl{LUT} remains a key challenge.
Existing approaches include noisy-critical-number-based optimization~\cite{PDDV2013} and the \gls{IBM}~\cite{LU2018,SWBW2020}. 
Although the \gls{IBM} provides excellent decoding performance, it generates node- and iteration-specific \glspl{LUT}. 
This approach is efficient for iteration-unrolled architectures for ultra-high throughput communications beyond $100$\,\unit{\giga\bit\per\second}~\cite{TG20}.
However, the approach is inefficient for non-unrolled fully node-parallel architectures that reuse the same processing units across iterations.
In such architectures, iteration-specific \glspl{LUT} require additional storage, multiplexing, and control logic. 
To address this limitation, we propose an iteration-agnostic \gls{LUT} design approach.
We use Bayesian optimization~\cite{M1974} with the \gls{FER} as the loss function to optimize the parameters of \gls{min-RCQ} decoders.
Furthermore, we introduce a search space reduction of the parameter space to enhance sample efficiency.
Bayesian optimization only needs access to the decoder parameters and the bit sequences at the transmitter and after decoding.
The Bayesian optimization needs no prior knowledge of the channel nor the modulation format.
Furthermore, the Bayesian optimization operates directly on the specific code rather than on an asymptotic cycle-free code as the \gls{IBM} does \cite{LU2018}.

Thus, we present a novel approach to determine the message values for a \gls{min-RCQ} decoder using Bayesian optimization that can be efficiently implemented on fully node-parallel architectures that process the decoding iterations sequentially. 
To improve sample efficiency, we reduce the parameter space by introducing a search area within it.
Furthermore, we show that a \gls{min-RCQ} decoder using our proposed \gls{LUT} design method can outperform a state-of-the-art \gls{NMS} decoder in both error-correction performance and implementation complexity, resulting in improved energy efficiency.

The code for creating the \glspl{LUT} within the \gls{min-RCQ} decoder is publicly available at~\cite{github-code}.

\section{Preliminaries}
\label{sec:Background}

\subsection{LDPC Codes}
Binary \gls{LDPC} codes~\cite{mackay} are defined by a sparse \gls{PCM} $\bm{H}\in\mathbb{F}_2^{M\times N}$. 
\gls{BP} decoders operate on the Tanner graph, which is a bipartite graph associated with the \gls{PCM}~\cite{RU2008}.
Two kinds of nodes exist in a Tanner graph: \glspl{CN} $\mathsf{c}_j,\, {j \in \{1, 2, \ldots, M\}}$, and \glspl{VN}~${\mathsf{v}_i, \,i \in \{1, 2, \ldots, N\}}$.
\gls{CN} $\mathsf{c}_j$ is connected to \gls{VN} $\mathsf{v}_i$, if $H_{j,i} = 1$.

\subsection{Conventional LDPC Decoder}
\gls{BP} decoders exchange messages between \glspl{CN} and \glspl{VN} along the Tanner graph edges connecting them.

In this paper, we use \gls{BP}-like algorithms in which messages are represented as \glspl{LLR}.
The \gls{LLR} of a binary random variable $X$ is denoted as:
\begin{align}
    L(X) = \ln{\left(\frac{P(X=0)}{P(X=1)}\right)}.
    \label{eq:likelihood}
\end{align}
Hence, the \gls{LLR} $L_i$ of a received code bit $y_i$ at the $i$-th position of the codeword of length $N$ is denoted as:
\begin{align}
    L_i = \ln{\left(\frac{P(Y_i=y_i|X_i=0)}{P(Y_i=y_i|X_i=1)}\right)}.
    \label{eq:likelihood_input}
\end{align}
The \glspl{LLR} $L_i$ are refined iteratively by exchanging extrinsic messages.
This is done based on the repetition code implied by the \glspl{VN} and the parity checks implied by \glspl{CN}.
We denote the neighborhood of \gls{VN} $\mathsf{v}_i$, i.e., the \glspl{CN} connected to $\mathsf{v}_i$, as $\mathcal{N}(i)$.
The estimates based on the repetition code are referred to as \gls{VN}-to-\gls{CN}-messages ${L_{i \rightarrow j}^{[\mathsf{v}]},\,\forall i\in\{1, 2, \ldots, N\},\,\, \forall j \in\mathcal{N}(i)}$. 
The messages based on the single parity checks are referred to as \gls{CN}-to-\gls{VN}-messages ${L_{i \leftarrow j}^{[\mathsf{c}]},\,\forall j\in\{1, 2, \ldots, M\},\,\, \forall i \in \mathcal{M}(j)}$.
Here, $\mathcal{M}(j)$ denotes the neighborhood of \gls{CN} $\mathsf{c}_j$, i.e., the \glspl{VN} connected to \gls{CN} $\mathsf{c}_j$, and $\mathcal{N}(i)$ the neighborhood of \gls{VN} $\mathsf{v}_i$.

The \gls{VN}-to-\gls{CN}-messages are initialized with the received \glspl{LLR} 
${L_{i \rightarrow j}^{[\mathsf{v}]} = L_i,\, \forall j \in\mathcal{N}(i)}$.
A common flavor of \gls{BP} algorithm is the \gls{NMS} algorithm~\cite{CDEFH2005}.
The \gls{NMS} algorithm computes the \gls{CN}-to-\gls{VN}-messages ${L_{i \leftarrow j}^{[\mathsf{c}]}}$ as:
\begin{align}
    L_{i \leftarrow j}^{[\mathsf{c}]} & = 
    \gamma
    \cdot
    \hspace*{-1mm}
    \left[
    \min_{i' \in \mathcal{M}(j)\setminus \{i\}}
    \hspace*{-1mm}
    \big\vert L_{i'\rightarrow j}^{[\mathsf{v}]} 
    \big\vert
    \right]
    \cdot
    \hspace*{-2mm}
    \prod_{i' \in \mathcal{M}(j)\setminus \{i\}} 
        \hspace*{-5mm}
        \textrm{sign}\left( L_{i'\rightarrow j}^{[\mathsf{v}]} \right).
    \label{eq:nms_cn2vn}
\end{align}
The scaling factor $\gamma$ is used to fine-tune the messages $L_{i \leftarrow j}^{[\mathsf{c}]}$ to better match the exact marginals.
The \gls{VN}-to-\gls{CN}-messages ${L_{i\rightarrow j}^{[\mathsf{v}]}}$ are given by:
\begin{align}
    L_{i\rightarrow j}^{[\mathsf{v}]} &= L_i + 
    \hspace*{-3mm}
    \sum_{j'\in \mathcal{N}(i)\setminus \{j\}} L_{i\leftarrow j'}^{[\mathsf{c}]} ,\quad \forall j \in \mathcal{N}(i)\; .
    \label{eq:nms_vn2cn}
\end{align}
These update steps are then iteratively repeated until the maximum number of allowed iterations is reached.
The refined estimates for the \gls{LLR} $\tilde{L}_{i}$ is finally calculated as:
\begin{align}
    \tilde{L}_{i} &= L_i + \sum_{j\in \mathcal{N}(i)} L_{i\leftarrow j}^{[\mathsf{c}]} \; .
    \label{eq:Estimate}
\end{align}
Finally, the decoded bits $\hat{x}_i$ are obtained by applying a hard decision 
$\hat{x}_i = \mathbbm{1}_{\{\tilde{L}_i \leq 0\}}$.

\gls{LDPC} decoders are typically implemented with finite precision. 
Using uniformly quantized messages with $64$\,levels, i.e., $6$\,\unit{\bit}, the \gls{NMS} algorithm approaches the \gls{FER} of floating-point \gls{NMS} at reasonable implementation complexity \cite{TG20}.
However, the complexity of the routing network becomes challenging for large codes in fully node-parallel architectures.

\subsection{Min-RCQ LDPC Decoder}
To enable larger codeword lengths, reducing the precision of the exchanged messages while maintaining error correction performance is crucial.
Decoders using the \gls{RCQ} approach meet these demands by exchanging low-precision messages that are mapped into a high-precision space for calculations~\cite{WWSB2020}.

We define the high-precision space $\mathcal{X}^\mathsf{h}$ with $2^{q_\mathsf{h}}$\,levels and the low-precision space $\mathcal{X}^\mathsf{l}$ with $2^{q_\mathsf{l}}<2^{q_\mathsf{h}}$\,levels as: 
\begin{align}
    \mathcal{X}^\mathsf{h} &=\left\{m \cdot \Delta x : m=\varphi(\bm{r}),\,\forall\bm r\in\mathbb{F}_2^{q_\mathsf{h}}\right\}\label{eq:back:Xh}\\
    \mathcal{X}^\mathsf{l} &= \left\{\sum_{\zeta = 0}^{n}\Delta x^\mathsf{l}(\zeta) : n=\varphi(\bm{u}),\,\forall\bm u\in\mathbb{F}_2^{q_\mathsf{l}}\right\}.\label{eq:back:Xl}
\end{align}
In \cref{eq:back:Xh,eq:back:Xl}, $\varphi(\cdot)$ denotes the mapping of a binary vector to its integer value in base-$10$.

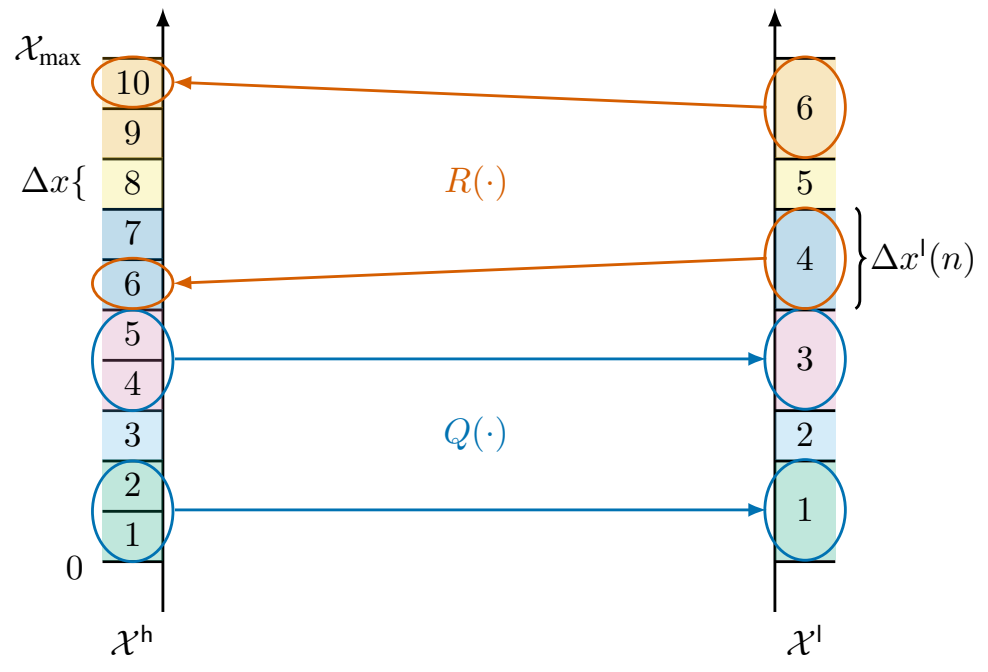
\begin{figure}
    \centering
    \resizebox{0.75\linewidth}{!}{\begin{tikzpicture}[>=latex,thick]
    \def\ydist{0.1cm}
    \def\xdist{0.3cm}
    \def\h{1.0cm}
    \node[] (xh) {$\mathcal{X}^\mathsf{h}$};
    \node[right =6.0cm of xh] (xl) {$\mathcal{X}^\mathsf{l}$};

    \node[above left = 0.2cm and 0cm of xh] (0) {$0$};
    \node[above left = 5.3cm and 0cm of xh] (0) {$\mathcal{X}_{\textrm{max}}$};

    \draw[->,color=OIblack] (xh.north) ++(+3mm,0) to ++(0,6.0cm);
    \draw[->,color=OIblack] (xl.north) ++(-3mm,0) to ++(0,6.0cm);


    \foreach \y in {0.5,1.0,...,5.5}
    {
        \draw[color=OIblack] (xh.north) ++(-3mm,\y cm) to ++(6mm,0);
    }

    \draw[color=OIblack] (xl.north) ++(-3mm,0.5cm) to ++(6mm,0);
    \node[above = 0.5cm of xh,draw,rectangle,draw opacity = 0.0 , minimum height=5mm, minimum width=6mm, fill=OIgreen,fill opacity = 0.25,text opacity=1] (z1h){$1$};
    \node[above = 1.0cm of xh,draw,rectangle,draw opacity = 0.0 , minimum height=5mm, minimum width=6mm, fill=OIgreen,fill opacity = 0.25,text opacity=1] (z1h){$2$};
    \node[above = 0.5cm of xl,draw,rectangle,draw opacity = 0.0 , minimum height=10mm, minimum width=6mm, fill=OIgreen,fill opacity = 0.25,text opacity=1] (z1l){$1$};
    \draw[color=OIblack] (xl.north) ++(-3mm,1.5cm) to ++(6mm,0);    
    \node[above = 1.5cm of xh,draw,rectangle,draw opacity = 0.0 , minimum height=5mm, minimum width=6mm, fill=OIcyan,fill opacity = 0.25,text opacity=1] (z2h){$3$};
    \node[above = 1.5cm of xl,draw,rectangle,draw opacity = 0.0 , minimum height=5mm, minimum width=6mm, fill=OIcyan,fill opacity = 0.25,text opacity=1] (z2l){$2$};
    \draw[color=OIblack] (xl.north) ++(-3mm,2.0cm) to ++(6mm,0);
    \node[above = 2.0cm of xh,draw,rectangle,draw opacity = 0.0 , minimum height=5mm, minimum width=6mm, fill=OIpink,fill opacity = 0.25,text opacity=1] (z3h){$4$};
    \node[above = 2.5cm of xh,draw,rectangle,draw opacity = 0.0 , minimum height=5mm, minimum width=6mm, fill=OIpink,fill opacity = 0.25,text opacity=1] (z3h){$5$};
    \node[above = 2.0cm of xl,draw,rectangle,draw opacity = 0.0 , minimum height=10mm, minimum width=6mm, fill=OIpink,fill opacity = 0.25,text opacity=1] (z3l){$3$};
    \draw[color=OIblack] (xl.north) ++(-3mm,3.0cm) to ++(6mm,0);    
    \node[above = 3.0cm of xh,draw,rectangle,draw opacity = 0.0 , minimum height=5mm, minimum width=6mm, fill=OIblue,fill opacity = 0.25,text opacity=1] (z4h){$6$};
    \node[above = 3.5cm of xh,draw,rectangle,draw opacity = 0.0 , minimum height=5mm, minimum width=6mm, fill=OIblue,fill opacity = 0.25,text opacity=1] (z4h){$7$};
    \node[above = 3.0cm of xl,draw,rectangle,draw opacity = 0.0 , minimum height=10mm, minimum width=6mm, fill=OIblue,fill opacity = 0.25,text opacity=1] (z4l){$4$};
    \draw[color=OIblack] (xl.north) ++(-3mm,4.0cm) to ++(6mm,0);
    \node[above = 4.0cm of xh,draw,rectangle,draw opacity = 0.0 , minimum height=5mm, minimum width=6mm, fill=OIyellow,fill opacity = 0.25,text opacity=1] (z5h){$8$};
    \node[above = 4.0cm of xl,draw,rectangle,draw opacity = 0.0 , minimum height=5mm, minimum width=6mm, fill=OIyellow,fill opacity = 0.25,text opacity=1] (z5l){$5$};
    \draw[color=OIblack] (xl.north) ++(-3mm,4.5cm) to ++(6mm,0);
    \node[above = 4.5cm of xh,draw,rectangle,draw opacity = 0.0 , minimum height=5mm, minimum width=6mm, fill=OIorange,fill opacity = 0.25,text opacity=1] (z6h){$9$};
    \node[above = 5.0cm of xh,draw,rectangle,draw opacity = 0.0 , minimum height=5mm, minimum width=6mm, fill=OIorange,fill opacity = 0.25,text opacity=1] (z6h){$10$};
    \node[above = 4.5cm of xl,draw,rectangle,draw opacity = 0.0 , minimum height=10mm, minimum width=6mm, fill=OIorange,fill opacity = 0.25,text opacity=1] (z6l){$6$};
    \draw[color=OIblack] (xl.north) ++(-3mm,5.5cm) to ++(6mm,0);

    \draw[OIblue] (z1l) ellipse (4mm and 5mm) node (el1){};
    \draw[OIblue] (z1h.south) ellipse (4mm and 5mm) node (eh1){};
    \draw[<-,color=OIblue] (el1.west)++(-2.5mm,0) to ++(-5.88cm,0.0cm);
    \node[left = 2.5cm of z2l] (l2) {\textcolor{OIblue}{$Q(\cdot)$}};
    \draw[OIblue] (z3l) ellipse (4mm and 5mm) node (el3){};
    \draw[OIblue] (z3h.south) ellipse (4mm and 5mm) node (eh3){};
    \draw[<-,color=OIblue] (el3.west)++(-2.5mm,0) to ++(-5.88cm,0.0cm);

    \draw[OIred] (z6l) ellipse (4mm and 5mm) node (el6){};
    \draw[OIred] (z6h)++(0,0mm) ellipse (4mm and 2.5mm) node (eh6){};
    \draw[->,color=OIred] (el6.west)++(-2.5mm,0) to ++(-5.9cm,0.25cm);
    \node[left = 2.5cm of z5l] (l5) {\textcolor{OIred}{$R(\cdot)$}};
    \draw[OIred] (z4l) ellipse (4mm and 5mm) node (el4){};
    \draw[OIred] (z4h)++(0,-5mm) ellipse (4mm and 2.5mm) node (eh4){};
    \draw[->,color=OIred] (el4.west)++(-2.5mm,0) to ++(-5.9cm,-0.25cm);

    \node[left =0.0cm of z5h] (dxh) {$\Delta x \{$};
    \draw [decorate,decoration = {brace}] (z4l)++(5mm,5mm) --  ++(0,-10mm);
    \node[right=2mm of z4l]{$\Delta x^\mathsf{l}(n)$};
\end{tikzpicture}}
    \caption{
        Exemplary quantization $Q(\cdot)$ from $\mathcal{X}^\mathsf{h}$ to $\mathcal{X}^\mathsf{l}$, and reconstruction $R(\cdot)$ from $\mathcal{X}^\mathsf{l}$ to $\mathcal{X}^\mathsf{h}$. 
        $Q(\cdot)$ maps the integers $m$ to $n$.
        $R(\cdot)$ maps vice versa.
        The mappings of $Q(\cdot)$ are highlighted by the background color, e.g., $\{1,2\}\mapsto1$.
        For $R(\cdot)$, we can choose for each element in $\mathcal{X}^\mathsf{l}$ any element of $\mathcal{X}^\mathsf{h}$ that is in $Q^{-1}(\mathcal{X}^\mathsf{l})$.
        Hence, $R(\cdot)$ maps, e.g., $4\mapsto6$, and $6\mapsto10$. 
    }
    \label{fig:back:mapping}
    \vspace*{-2mm}
\end{figure}

\cref{fig:back:mapping} shows exemplary mappings between~$\mathcal{X}^\mathsf{h}$ and $\mathcal{X}^\mathsf{l}$.
Observe that~$\mathcal{X}^\mathsf{h}$ has uniform step size $\Delta x$, whereas~$\mathcal{X}^\mathsf{l}$ has varying step sizes~$\Delta x^\mathsf{l}(n)$.
We denote the mapping from $\mathcal{X}^\mathsf{h}$ to $\mathcal{X}^\mathsf{l}$ as quantization $Q(\cdot)$ and mapping from $\mathcal{X}^\mathsf{l}$ to $\mathcal{X}^\mathsf{h}$ as reconstruction $R(\cdot)$.
The quantization function $Q(\cdot)$ can map several values to a single value in $\mathcal{X}^\mathsf{l}$.
The reconstruction $R(\cdot)$ maps a value in $\mathcal{X}^\mathsf{l}$ to a single value of $\mathcal{X}^\mathsf{h}$, that has to be from the pre-image $Q^{-1}(\mathcal{X}^\mathsf{l})$.
$Q(\cdot)$ and $R(\cdot)$ can be implemented as an integer-to-integer mapping using \glspl{LUT}.

\begin{figure}
    \centering
    \resizebox{0.8\linewidth}{!}{\begin{tikzpicture}[>=latex,thick]
    \def\ydist{0.1cm}
    \def\xdist{0.3cm}
    \def\h{1.0cm}
    \node[] (L1) {$L_{i}$};

    \node[right= 1.9*\xdist of L1,draw,rectangle,rounded corners, minimum height=0.5*\h, minimum width=1.1*\h] (rec1) {$Q_{\Delta x}(\cdot)$}; 
    \node[below= 2.5*\ydist of rec1,draw,rectangle,rounded corners, minimum height=0.5*\h, minimum width=1.1*\h] (rec2) {$R_{i \rightarrow j_1^{i}}^{[\mathsf{v}]}(\cdot)$};
    \node[below= 1*\ydist of rec2,draw,rectangle,rounded corners, minimum height=0.5*\h, minimum width=1.1*\h] (rec3) {$R_{i \rightarrow j_2^{i}}^{[\mathsf{v}]}(\cdot)$};
    \node[below right= 5*\ydist and -15.5mm of rec3,draw,rectangle,rounded corners, minimum height=0.5*\h, minimum width=1.1*\h] (rec4) {$R_{i \rightarrow j_{d_\mathsf{v}}^{i}}^{[\mathsf{v}]}\hspace*{-1.1mm}(\cdot)$};

    \node[left= 4mm of rec2] (L2) {$L_{i \leftarrow j_{1}^{i}}^{[\mathsf{c}]}$};
    \node[left= 4mm of rec3] (L3) {$L_{i \leftarrow j_{2}^{i}}^{[\mathsf{c}]}$};
    \node[left= 3mm of rec4] (L4) {$L_{i \leftarrow j_{d_\mathsf{v}}^{i}}^{[\mathsf{c}]}$};

    \draw[->,color=OIpink] (L1) to (rec1.west);
    \draw[->,color=OIred] (L2) to (rec2.west);
    \draw[->,color=OIred] (L3) to (rec3.west);
    \draw[<-,color=OIred] (rec4.west) to ++(-4mm,0);

    \node[below right=-0.5*\h and 2.5*\xdist of rec1,draw,rectangle,rounded corners, minimum height=1.0*\h] (sum) {\Large$\Sigma$};
    \node[right=4.5*\xdist of rec2,draw,circle, inner sep=0pt,radius=0.1cm] (m1) {$-$};
    \node[right=5.5*\xdist of rec3,draw,circle, inner sep=0pt,radius=0.1cm] (m2) {$-$};
    \node[right=6.5*\xdist of rec4,draw,circle, inner sep=0pt,radius=0.1cm] (m3) {$-$};

    \draw[->,color=OIgreen] (rec1) to (\tikztostart -| sum.west);
    \draw[->,color=OIgreen] (rec2) to (m1.west);
    \draw[->,color=OIgreen] (rec3) to (m2.west);
    \draw[->,color=OIgreen] (rec4) to (m3.west);

    \node[right=1*\xdist-2mm of rec2,draw,circle,inner sep=1, fill=OIgreen,color=OIgreen] (b1) {};
    \node[right=1*\xdist-1mm of rec3,draw,circle,inner sep=1, fill=OIgreen,color=OIgreen] (b2) {};
    \node[right=1*\xdist+0.1mm of rec4,draw,circle,inner sep=1, fill=OIgreen,color=OIgreen] (b3) {};

    \node[above=3.8mm of m1,draw,circle,inner sep=1, fill=OIblue,color=OIblue] (c1) {};
    \node[above=13.3mm of m2,draw,circle,inner sep=1, fill=OIblue,color=OIblue] (c2) {};
    \node[above=27.0mm of m3,draw,circle,inner sep=1, fill=OIblue,color=OIblue] (c3) {};
    
    \node[below right = 1mm   and 5mm of b2] () {$\vdots$};
    \node[below right = 0.5mm and 2.2cm of L1] () {$\cdot$};
    \node[below right = 1.3mm and 2.2cm of L1] () {$\cdot$};
    \node[below right = 2.1mm and 2.2cm of L1] () {$\cdot$};

     \draw[->,color=OIgreen] (b1) |- ++(5mm  , 7mm);
     \draw[->,color=OIgreen] (b2) |- ++(4mm  ,14.5mm);
     \draw[->,color=OIgreen] (b3) |- ++(3.0mm,25mm);

     \draw[->,color=OIblue] (c1) to (m1.north);
     \draw[->,color=OIblue] (c2) to (m2.north);
     \draw[->,color=OIblue] (c3) to (m3.north);

    \node[right= 3.0*\xdist of m1,draw,rectangle,rounded corners, minimum height=0.5*\h] (Q1) {$Q_{i \rightarrow j_1^{i}}^{[\mathsf{v}]}(\cdot)$};
    \node[below= 1 *\ydist of Q1,draw,rectangle,rounded corners, minimum height=0.5*\h] (Q2) {$Q_{i \rightarrow j_2^{i}}^{[\mathsf{v}]}(\cdot)$};
    \node[below right= 5*\ydist and -15.7mm of Q2,draw,rectangle,rounded corners, minimum height=0.5*\h] (Q3) {$Q_{i \rightarrow j_{d_\mathsf{v}}^{i}}^{[\mathsf{v}]}\hspace*{-1.1mm}(\cdot)$};

    \node[right=1.5*\xdist of Q1] (Lo1) {$L_{i \rightarrow j_1^{i}}^{[\mathsf{v}]}$};
    \node[right=1.5*\xdist of Q2] (Lo2) {$L_{i \rightarrow j_2^{i}}^{[\mathsf{v}]}$};
    \node[right=1.5*\xdist of Q3] (Lo3) {$L_{i \rightarrow j_{d_\mathsf{v}}^{i}}^{[\mathsf{v}]}$};
    \node[above=-0.1cm of Lo1] (Ltd) {$\tilde{L}_{i}$};

    \draw[<-,color=OIblue] (Ltd.west) to (\tikztostart -| sum.east);
    
    \draw[->,color=OIblue] (m1) -- (Q1);
    \draw[->,color=OIblue] (m2) -- (Q2);
    \draw[->,color=OIblue] (m3) -- (Q3);
    
    \draw[->,color=OIred] (Q1) -- (Lo1);
    \draw[->,color=OIred] (Q2) -- (Lo2);
    \draw[->,color=OIred] (Q3) -- (Lo3);

    \node[draw,dash dot,color=OIred] at (0.1,-4.0) {$q_\mathsf{l}$ bit};
    \node[draw,dash dot,color=OIgreen] at (3.1,-4.0) {$q_\mathsf{h}$ bit};
    \node[draw,dash dot,color=OIblue] at (5.5, 0.4) {$q_\mathsf{h}+\left\lceil\log_2(d_\mathrm{v}+1)\right\rceil$ bit};
    \node[draw,dash dot,color=OIred] at (7.2,-4.0) {$q_\mathsf{l}$ bit};
    \node[draw,dash dot,color=OIpink] at (0.5,0.4) {$\mathbb{R}$};

\end{tikzpicture}}
    \caption{
        Fully parallel architecture of a \gls{min-RCQ} \gls{VN} update unit.
    }
    \label{fig:back:vnu_rcq}
    \vspace*{-\PicSpace}
\end{figure}
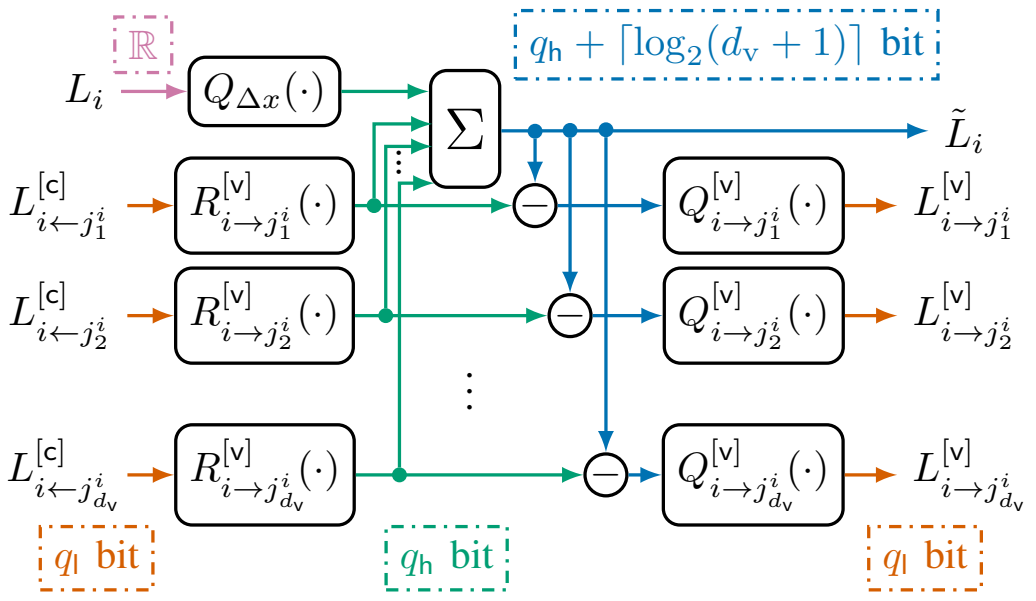

The \gls{CN} updates of the \gls{min-RCQ} decoder follow the \gls{MS} update rule \cref{eq:nms_cn2vn} with ${\gamma=1}$.
The major difference between an \gls{MS}-based and a \gls{min-RCQ} decoder is the \gls{VN} update.
The \gls{VN} updates of the \gls{min-RCQ} decoder are similar to the \gls{MS} update rule \cref{eq:nms_vn2cn}, but incorporate reconstruction and quantization functions.
The \gls{min-RCQ} \gls{VN} update unit is shown in \cref{fig:back:vnu_rcq}.
The channel \glspl{LLR} $L_i\in\mathbb{R}$ are quantized by a uniform quantizer $Q_{\Delta x}(\cdot)$ with precision $q_\mathsf{h}$ bits. 
The \gls{CN}-to-\gls{VN}-messages ${L_{i \leftarrow j}^{[\mathsf{c}]}\in\mathcal{X}^\mathsf{l}}$ are mapped with the reconstruction functions 
${R_{i\rightarrow j}^{[\mathsf{v}]}(\cdot)}$ to the high-precision space $\mathcal{X}^\mathsf{h}$.
After the calculation of the update step \cref{eq:nms_vn2cn}, the non-uniform quantizers ${Q_{i\rightarrow j}^{[\mathsf{v}]}(\cdot)}$ map the \gls{VN}-to-\gls{CN}-messages ${L_{i\rightarrow j}^{[\mathsf{v}]}}$ to the low-precision space $\mathcal{X}^\mathsf{l}$.
The quantizers and reconstructions can vary from edge to edge in the Tanner graph, whereas the uniform \gls{LLR} quantizers $Q_{\Delta x}(\cdot)$ are shared throughout the decoder.
Hence, the \gls{VN}-to-\gls{CN} update rule from \cref{eq:nms_vn2cn} can be re-formulated as:
\begin{align}
    L_{i\rightarrow j}^{[\mathsf{v}]}\hspace*{-0.5mm} &= \hspace*{-0.5mm} Q_{i\rightarrow j}^{[\mathsf{v}]}
    \hspace*{-1mm}
    \left( 
    \hspace*{-1mm}
    Q_{\Delta x}(L_i) + 
    \hspace*{-6mm}
    \sum_{j'\in \mathcal{N}(i)\setminus \{j\}} 
    \hspace*{-6mm}
    R_{i\rightarrow j'}^{[\mathsf{v}]}\hspace*{-1mm}\left(L_{i\leftarrow j'}^{[\mathsf{c}]}\hspace*{-1mm}\right) 
    \hspace*{-1.5mm}
    \right)
    \hspace*{-1mm}
    ,\forall j \hspace*{-1mm} \in \hspace*{-0.5mm} \mathcal{N}(i).
    \label{eq:rcq_vn2cn}
\end{align}

\subsection{Bayesian Optimization}
Bayesian optimization is a suitable global optimization technique for a minimization problem $\min_{\bm\theta\in\mathcal{S}} F(\bm\theta)$, where the evaluation of the loss function $F(\bm\theta)$ is time-consuming and/or gradient descent techniques~\cite{M1974, JSW1998} are not feasible.

We define $\mathcal{S}$ as the parameter space, and $\bm{\theta}\in\mathcal{S}$ as parameter vectors.
A data set containing $D$ parameter vectors and the corresponding losses is denoted as ${\mathcal{A}_D = \{(\bm\theta_i,F(\bm\theta_i))\}_{i=1}^D}$.
We assume that the losses ${F(\bm\theta_i)=U(\bm{\theta}_i)+e,\,e\sim\mathcal{N}(0,\sigma^2)}$, are noisy observations of an unknown true loss function $U(\bm{\theta}_i)$.
Since $U(\bm{\theta})$ is unknown and $F(\bm\theta)$ is expensive to evaluate, we want to reduce the number of samples drawn from $F(\bm\theta)$ by interpolating the data set $\mathcal{A}_D$ to find promising regions within $\mathcal{S}$ to sample from.

A common regression for such a task is \gls{GPR}.
A Gaussian process $\mathcal{GP}(m(\bm\theta),k(\bm{\theta},\bm{\theta}^\prime))$ is a collection of random variables, any finite number of which have a joint Gaussian distribution.
The mean function is defined as $m(\bm\theta) = \mathbb{E}\{F(\bm\theta)\}$ and the covariance function as~${k(\bm{\theta},\bm{\theta}^\prime)=\mathbb{E}\{(F(\bm\theta)-m(\bm\theta))\,(F(\bm\theta^\prime)-m(\bm\theta^\prime))\}}$~\cite{RW2006}.
We denote the covariance matrix of the database $\mathcal{A}_D$ as $\bm{K}$.
The regression of the loss function for a new point $\bm{\theta}^\prime$ is given 
by~${\tilde{F}(\bm\theta^\prime) = \mathcal{GP}(m(\bm\theta^\prime|\mathcal{A}_D),k(\bm{\theta}^\prime|\mathcal{A}_D}))$~\cite{RW2006}:
\begin{align}
    m(\bm\theta^\prime|\mathcal{A}_D) &= \bm{k}(\bm\theta^\prime)^\mathsf{T}(\bm{K}+\sigma^2I)^{-1}\bm{F}(\bm\theta^\ast)\label{background:eq:predition}\\
    \mathbb{V}(\bm{\theta}^\prime|\mathcal{A}_D) &= k(\bm\theta^\prime,\bm{\theta}^\prime) 
                                           - \bm{k}(\bm\theta^\prime)^\mathsf{T}
                                             (\bm{K}+\sigma^2\bm{I})^{-1}
                                             \bm{k}(\bm\theta^\prime).
\end{align}
$\bm{F}(\bm\theta^\ast)$ contains all losses of $\mathcal{A}_D$,
$m(\bm\theta^\prime|\mathcal{A}_D)$ denotes the mean of the regression at $\bm\theta^\prime$,
$\mathbb{V}(\bm{\theta}^\prime|\mathcal{A}_D)$ denotes the variance of the regression at $\bm\theta^\prime$, and 
$\bm{k}(\bm\theta^\prime)$ contains the covariances between the point to interpolate $\bm\theta^\prime$ and the data set $\mathcal{A}_D$.
The mean function and covariance function of the \gls{GP} are adapted to match $\mathcal{A}_{D}$ using Alg.~2.1 of~\cite[p.~19]{RW2006}.

At the beginning of the Bayesian optimization, the dataset $\mathcal{A}_D$ is created by randomly sampling the parameter vectors $\bm\theta_i$ and evaluating them using the expensive $F(\bm\theta_i)$.
We define the best loss in the data set as~${F_{\textrm{min}} = \min_{\bm\theta^\ast}F(\bm\theta^\ast),\,\forall(\bm\theta^\ast, F(\bm\theta^\ast))\in \mathcal{A}_D}$.
In each iteration $z$, the best loss of the dataset $\mathcal{A}_{D+z-1}$ is re-evaluated, and a new parameter vector $\bm\theta_z$ is evaluated using the expensive $F(\bm\theta_z)$.
The result is added to the data~set~${\mathcal{A}_{D+z} = \mathcal{A}_{D+z-1}\cup\{\left(\bm\theta_{z},F(\bm\theta_{z})\right)\}}$.
The mean function and covariance function of the \gls{GP} are adapted in each iteration to match $\mathcal{A}_{D+z}$.
Since evaluating the \gls{GPR} is much cheaper in complexity and time than evaluating $F(\bm\theta_{z})$, the \gls{GPR} is used to find an optimal candidate to evaluate in each iteration $\bm\theta_{z}$.
With the \gls{GPR} (providing the statistics of the regression) and the best loss, we can calculate the likelihood of $\bm\theta^\prime\in\mathcal{S}$ to perform better than the best data set loss, i.e., $F(\bm\theta^\prime)<F_{\textrm{min}}$.
With $\mathrm{H}(x)$ as the Heaviside step function, the probability of improvement is defined as \gls{EI}~\cite{JSW1998}:
\begin{align}
    \mathrm{EI}(\bm\theta^\prime) = \mathbb{E}\{\mathrm{H}(F_{\textrm{min}}-\tilde{F}(\bm\theta^\prime))\}.
\end{align}
\gls{EI} is used to efficiently find the parameter to evaluate next~${\bm\theta_{z} = \arg\max_{\bm\theta^\prime\in\mathcal{S}}\mathrm{EI}(\bm\theta^\prime)}$.
This is based on the assumption that the maximization of $\mathrm{EI}(\bm\theta^\prime)$, i.e., the \gls{GPR} $\tilde{F}(\bm\theta^\prime)$, maximizes the improvement for the true loss function $U(\bm\theta^\prime)$.

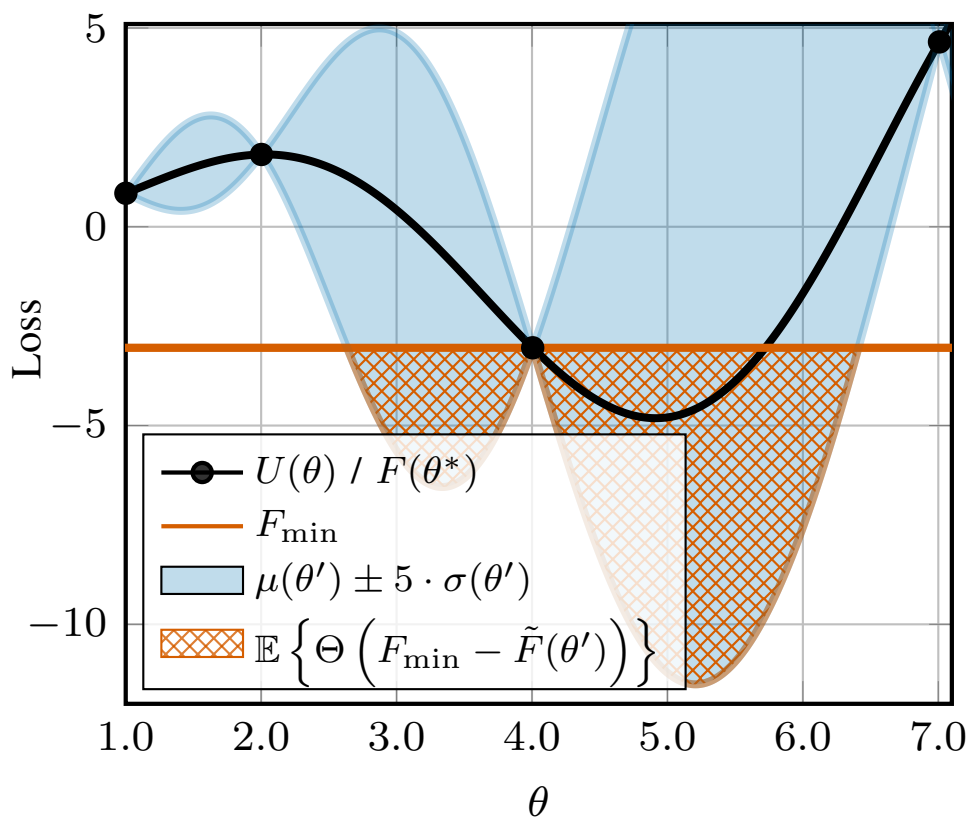
\begin{figure}[t]
    \centering
    \pgfplotsset{}
\usepgfplotslibrary{fillbetween}
\usetikzlibrary{patterns}

\begin{tikzpicture}[every node/.append style={font=\footnotesize}]
    \def\lwidth{1.5}
    \def\marksz{1.5pt}
    \def\markszl{3pt}

    \begin{axis}[
        width = 0.8\columnwidth,
        xscale=1,
        xlabel style = {align=center},
        xlabel=$\theta$,
        ylabel={Loss},
        y label style={at={(axis description cs:-0.12,0.45)},anchor=west},
        xtick = {1.0,2.0,3.0,4.0,5.0,6.0,7.0,8.0},
        xticklabels={$1.0$,$2.0$,$3.0$,$4.0$,$5.0$,$6.0$,$7.0$,$8.0$},
        grid=major,
        legend cell align={left},
        legend style={
            at={(0.02,0.02)},
            anchor=south west,
            fill opacity = .8,
            draw opacity = 1, 
            text opacity = 1,
            nodes={scale=0.9, transform shape}
        },
        xmin=1.0,xmax=7.1,
        ymin=-12,ymax=5.1,
        axis line style=thick,
        tick label style={/pgf/number format/fixed},
        legend image post style={scale=0.9},
        ]

        \addplot[mark=none, color=black,line width=\lwidth,forget plot] table [x = theta, y = L]{./figures/GP_Demo.txt}; 
        \addplot[only marks, mark=*, mark color=black,forget plot] table [x = theta, y = L]{./figures/GP_Demo_Samples.txt}; 
        
        \addplot[mark=none, color=OIred,line width=\lwidth,forget plot] table [x = theta, y = L_min]{./figures/GP_Demo.txt}; 

        \addplot+[name path=Low,mark=none, color=OIblue,line width=\lwidth, opacity= 0.25, forget plot] table [x = theta, y = Low]{./figures/GP_Demo.txt};
        \addplot+[name path=High,mark=none, color=OIblue,line width=\lwidth, opacity= 0.25, forget plot] table [x = theta, y = High]{./figures/GP_Demo.txt};
        \addplot [color=OIblue, opacity= 0.25,forget plot] fill between [of=Low and High];

        \addplot+[name path=ALow,mark=none, color=OIred,line width=\lwidth, opacity= 0.5, forget plot] table [x = theta, y = Area_below]{./figures/GP_Demo.txt};
        \addplot+[name path=AHigh,mark=none, color=OIred,line width=\lwidth, opacity= 0.5, forget plot] table [x = theta, y = L_min]{./figures/GP_Demo.txt};
        \addplot [pattern=crosshatch,
    pattern color=OIred,forget plot] fill between [of=ALow and AHigh];
        
        \addlegendimage{black, thick, mark=*, mark options={solid}} \addlegendentry{\footnotesize $U(\theta)$ / $F(\theta^\ast)$}
        \addlegendimage{OIred, thick} \addlegendentry{\footnotesize $F_\mathrm{min}$}
        \addlegendimage{fill=OIblue, fill opacity=0.25, draw=none, area legend} \addlegendentry{\footnotesize $\mu(\theta^\prime)\pm 5\cdot \sigma(\theta^\prime)$}
        \addlegendimage{pattern=crosshatch, pattern color=OIred, draw=OIred, area legend} \addlegendentry{\footnotesize $\mathbb{E}\left\{ \Theta\left( F_{\min}-\tilde{F}(\theta^\prime) \right) \right\}$}

\end{axis}
\end{tikzpicture}
    \caption{From the one dimensional loss function $U(\theta)$ noisy samples $F(\theta^\ast)$ are drawn. Using the samples, the \gls{GPR} $\tilde{F}(\theta^\prime)$ is adapted to approximate $U(\theta)$. The $5\sigma$ interval of $\tilde{F}(\theta^\prime)$ is plotted (blue). The region between the lowest observed loss $F_\mathrm{min}$ and the $5\sigma$ interval below $F_\mathrm{min}$ corresponds to $\mathrm{EI}(\theta^\prime)$ (red crosses).}
    \label{fig:gp_demo}
    \vspace*{-\PicSpace}
\end{figure}

\cref{fig:gp_demo} shows an example for a one-dimensional optimization problem. 
The \gls{EI} corresponds to the area of the $5\sigma$ interval below $F_\mathrm{min}$ (red stripes).
The \gls{EI} has the advantage of finding areas with high likelihood of containing parameters that reach lower losses than the currently observed, thereby enhancing the sample efficiency. 

In this work, we use the Bayesian optimizer with multivariate \glspl{GP} of scikit-optimize~\cite{scikit}.

\newpage
\section{Algorithm}
\label{sec:Algorithm}

In this section, we propose a novel \gls{LUT} design algorithm that enables the design and efficient implementation of \gls{min-RCQ} decoders on fully node-parallel \gls{LDPC} decoder architectures using Bayesian optimization.
Bayesian optimization is a suitable optimization method, since \gls{FER} simulations are time-consuming and no gradients are available.

\label{sec:Algorithm:LUT_Design}
{
    \begin{algorithm}[h]
        \DontPrintSemicolon
        \caption{\gls{min-RCQ} Bayesian optimization}
            Sample $\mathcal{A}_D$\;
            \For{$z = 1$ to $N_\textrm{max}$}{
                $F_{\textrm{min}} = \min_{\bm\theta^\ast}F(\bm\theta^\ast),\,\forall(\bm\theta^\ast,F(\bm\theta^\ast))\in \mathcal{A}_{D+z-1}$\;
                $\tilde{F}(\bm\theta^\prime) = \mathcal{GP}(m(\bm\theta^\prime|\mathcal{A}_{D+z-1}),k(\bm\theta^\prime|\mathcal{A}_{D+z-1}))$\;
                $\bm\theta_{z} = \arg\max_{\bm\theta^\prime\in\mathcal{S}} \mathbb{E}\left\{ \mathrm{H}\left( F_{\mathsf{min}}-\tilde{F}(\bm\theta^\prime) \right) \right\}$\;
                $\Delta x \leftarrow 
                    \mathrm{Step\_Size}(\bm\theta_{z}
                    [0])$\;
                $P_\mathsf{os} = 1$\;
                \For{ $d_{\mathsf{c}_z} = d_{\mathsf{c}_\textrm{min}}$ to $d_{\mathsf{c}_\textrm{max}}$}
                {
                    $\mathcal{Q} = \{Q_{i\rightarrow j}^{[\mathsf{v}]}:\sum_{i=1}^NH_{j,i}=d_{\mathsf{c}_z}\}$\; 
                    $\mathcal{R} = \{R_{i\rightarrow j}^{[\mathsf{v}]}:\sum_{i=1}^NH_{j,i}=d_{\mathsf{c}_z}\}$\; 
                    $\mathcal{Q},\mathcal{R} \leftarrow 
                        \mathrm{Sym\_LUT}(
                            \bm\theta_{z}[
                                P_\mathsf{os}:P_\mathsf{os}+2^{q_\mathsf{l}-1}-1
                            ]
                        )$\;
                    $P_\mathsf{os}=P_\mathsf{os}+2^{q_\mathsf{l}-1}-1$\;
                }
                $F(\bm\theta_{z})=\mathrm{FER}(\bm\theta_{z})$\;
                $\mathcal{A}_{D+z} = \mathcal{A}_{D+z-1}\cup\left(\bm\theta_{z},F(\bm\theta_{z})\right)$\;
            }
            \Return $\arg\min_{\bm\theta^\ast}\hspace*{-1mm}F(\bm\theta^\ast),\,\forall(\bm\theta^\ast,F(\bm\theta^\ast))\in \mathcal{A}_{D+N_\textrm{max}}$\;
        \label{alg:Algo1}
    \end{algorithm}
}

In \cref{alg:Algo1}, the Bayesian optimization of the \gls{min-RCQ} decoder parameter $\bm\theta$ is described.
At the beginning of the optimization, $D$ parameter vectors $\{\bm\theta_i\}_{i=1}^D$ are randomly sampled and evaluated.
The parameter vectors $\bm\theta_i$ contain the step size~$\Delta_x$ of~$\mathcal{X}^\mathsf{h}$ and the symmetrical \glspl{LUT}. 

We use the \gls{FER} as the loss function, i.e., $F(\bm\theta_i) = \textrm{FER}(\bm\theta_i)$.
The \gls{FER} is evaluated using Monte Carlo simulations with $C_\mathsf{w}$ codewords.
The \gls{FER} is based on the hard decision output of the \gls{min-RCQ} decoder after $I$ decoding iterations.

In each optimization iteration, the \gls{GPR} $\tilde{F}(\bm\theta^\prime)$ is adapted to $\mathcal{A}_{D+z-1}$.
A new parameter vector $\bm\theta_{z}$ is chosen based on the maximum \gls{EI}.
Based on this, $\bm\theta_{z}$ is split into the quantizer step size $\Delta_x$ and the symmetrical \glspl{LUT}. 
The \glspl{LUT} are assigned to $Q_{i\rightarrow j}^{[\mathsf{v}]}$ and $R_{i\rightarrow j}^{[\mathsf{v}]}$.
The min-RCQ decoder is then evaluated to obtain $\textrm{FER}(\bm\theta_{z})$. 
The resulting tuple $(\bm\theta_{z},\,\textrm{FER}(\bm\theta_{z}))$ is added to the dataset to create the new dataset $\mathcal{A}_{D+z}$.
In each optimization iteration, $\mathcal{A}_{D+z-1}$ is evaluated to find the smallest \gls{FER} $F_\textrm{min}$.
Finally, the parameter vector corresponding to $F_\textrm{min}$ is returned. 

Each edge in the Tanner graph can have distinct quantizers $Q_{i\rightarrow j}^{[\mathsf{v}]}$ and reconstructions $R_{i\rightarrow j}^{[\mathsf{v}]}$.
However, each $Q_{i\rightarrow j}^{[\mathsf{v}]}$ interfacing the same \gls{CN} should have the same \gls{LUT}, since otherwise the minimum search in $\mathcal{X}^\mathsf{l}$ produces different results than a minimum search in $\mathcal{X}^\mathsf{h}$.
Therefore, we grouped $Q_{i\rightarrow j}^{[\mathsf{v}]}$ and $R_{i\rightarrow j}^{[\mathsf{v}]}$ such that all edges interfacing a \gls{CN} of the same degree have the same \gls{LUT}.

\begin{figure}[t]
    \centering
    \resizebox{0.75\linewidth}{!}{\begin{tikzpicture}[>=latex]
    \begin{axis}[
        axis lines=middle,
        xlabel = {$n$},
        xlabel style={at={(axis cs:4,-3.0)}, anchor=north},
        ylabel={$m$},
        ylabel style={at={(axis cs:-1,18.5)}, anchor=north},
        xtick = {1,2,3,4,5,6,7},
        xticklabels ={$1$,$2$,$3$,$4$,$5$,$6$,$7$},
        ytick = {1,5,10,15,20,25,30},
        yticklabels ={$1$,$5$,$10$,$15$,$20$,
                      $25$,$30$},
        ymin=-0.5, ymax=31.5,
        xmin=-0.5, xmax=7.5,
        domain=-4:4,
        samples=300,
        thick,
        legend style={at={(axis cs:7.1,23.0)},anchor=north west}
    ]

    \addplot[mark=square*,mark options={solid,fill=OIred}, color=OIred, very thick] coordinates {(1,1) (2,2) (3,3) (4,4) (5,5) (6,6) (7,7)};
    \addplot[mark=triangle*,mark options={solid,fill=OIred}, color=OIred, very thick] coordinates {(1,25) (2,26) (3,27) (4,28) (5,29) (6,30) (7,31)};

    \addplot[mark=*,mark options={solid,fill=OIgreen}, OIgreen, very thick, dotted] coordinates {(1,4) (2,8) (3,12) (4,16) (5,20) (6,24) (7,28)};
    \addplot[mark=square*,mark options={solid,fill=OIgreen}, OIgreen, very thick] coordinates {(1,1) (2,4) (3,8) (4,12) (5,16) (6,20) (7,24)};
    \addplot[mark=triangle*,mark options={solid,fill=OIgreen}, OIgreen, very thick] coordinates {(1,8) (2,12) (3,16) (4,20) (5,24) (6,28) (7,31)};

    \addlegendentry{ $R_{\textrm{LB}}(n)$}
    \addlegendentry{ $R_{\textrm{UB}}(n)$}
    
    \addlegendentry{ $R_{s,\textrm{m}}(n)$}
    \addlegendentry{ $R_{\textrm{s,LB}}(n)$}
    \addlegendentry{ $R_{\textrm{s,UB}}(n)$}

    
    \end{axis}

    \draw [<->,black,line width=0.5mm] (2.14,1.6) -- (2.14,2.2);
    \node at (1.83,1.73) {$F$};
        
\end{tikzpicture}}
    \caption{
        First quadrant of the reconstruction functions boundaries for $\mathcal{X}^\mathsf{h}$ precision of $6$\,\unit{\bit} and $\mathcal{X}^\mathsf{l}$ of $4$\,\unit{\bit}.
        $m$ is the index of the value in $\mathcal{X}^\mathsf{h}$, whereas $n$ is the index of the value in $\mathcal{X}^\mathsf{l}$.
        The orange lines are the bounds given by the restriction of monotonic growth.
        The green lines define the bounds of the subspace defined in \cref{sec:Algorithm:LUT_Design}.
    }
    \label{fig:space}
    \vspace*{-\PicSpace}
\end{figure}
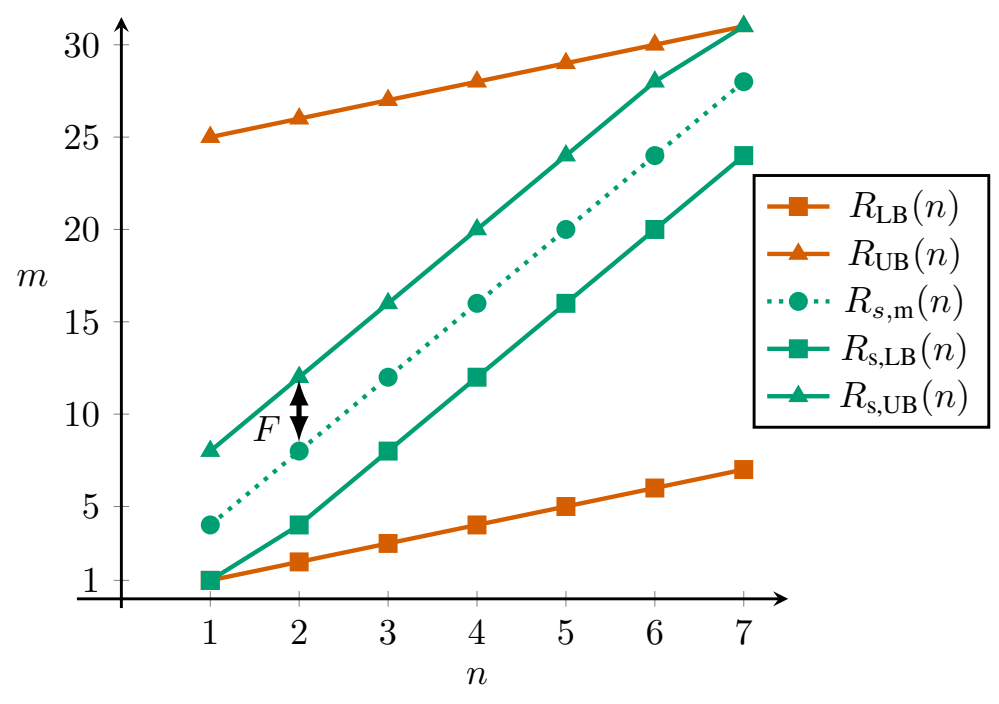

It is known that the reconstruction values should be monotonic \cite{PDDV2013}.
Furthermore, we restrict the \glspl{LUT} to be symmetric and contain $\pm0$.
Hence, $R_{i\rightarrow j}^{[\mathsf{v}]}$ and $Q_{i\rightarrow j}^{[\mathsf{v}]}$ can be fully described by the first quadrant of the function.
Furthermore, the step sizes of $\mathcal{X}^\mathsf{l}$ are defined by the pre-image of $Q_{i\rightarrow j}^{[\mathsf{v}]}$.
For simplicity, we omit the step sizes and describe the functions as integer-to-integer mappings.
With this, we define the lowest trajectory $R^i_{\textrm{LB}}$ and highest trajectory $R^i_{\textrm{UB}}$ of~$R_{i\rightarrow j}^{[\mathsf{v}]}$ as:
\begin{align*}
    |R_{\textrm{LB}}(n)| &= |n|\\
    |R_{\textrm{UB}}(n)| &= 2^{q_\mathsf{h}-1}-1 - (2^{q_\mathsf{l}-1}-1) + |n|\\
    \forall i &\in \{1, 2, \ldots, N\}.
\end{align*}
We have evaluated the \gls{FER} for all possible parameter vectors~$\bm\theta$ combinations for several channel codes, $\mathcal{X}^\mathsf{h}$ precision $q_\mathsf{h}$, and $\mathcal{X}^\mathsf{l}$ precision $q_\mathsf{l}$.
We observed that all parameter vectors~$\bm\theta$ with 
$\textrm{FER}(\bm\theta)\leq 1.1\cdot F_{\textrm{min}}$ 
were confined in a region whose boundaries are defined by:
\begin{align*}
    F &= 2^{q_\mathsf{h}-1}/2^{q_\mathsf{l}-1}\\
    |R_{s,\textrm{m}}(n)| &= F\cdot |n|\\
    |R_{\textrm{s,LB}}(n)| &= \max\{|R_{s,\textrm{m}}(n)|-F,\,|R_{\textrm{LB}}(n)|\}\\
    |R_{\textrm{s,UB}}(n)| &= \min\{|R_{s,\textrm{m}}(n)|+F,\,|R_{\textrm{UB}}(n)|\}.
\end{align*}
The function $R_{s,\textrm{m}}(\cdot)$ defines the mapping of an \gls{MS} decoder that spans across the space using equidistant steps.
The functions $R_{\textrm{s,LB}}(\cdot)$ and $R_{\textrm{s,UB}}(\cdot)$ define the boundaries of the reduced search space. They are created by offsetting $R_{s,\textrm{m}}(\cdot)$ by $F$ and bounding the region by the original space boundaries $R_{\textrm{LB}}(\cdot)$ and $R_{\textrm{UB}}(\cdot)$.
The search space is reasonable, since the areas outside the subspace cover only a fraction of the range from $0$ to $(2^{q_\mathsf{h}}-1)\cdot\Delta x$ or have large jumps in between.
An example with $q_\mathsf{l}$ of $4$\,\unit{\bit} and $q_\mathsf{h}$ of $6$\,\unit{\bit} for the lowest~$R_{\textrm{LB}}(n)$, highest trajectory $R_{\textrm{UB}}(n)$ and of the boundaries of the reduced search area $R_{\textrm{s,LB}}(n)$ and $R_{\textrm{s,UB}}(n)$ is shown in~\cref{fig:space}.

\section{Decoder Architecture}
\label{sec:arch}

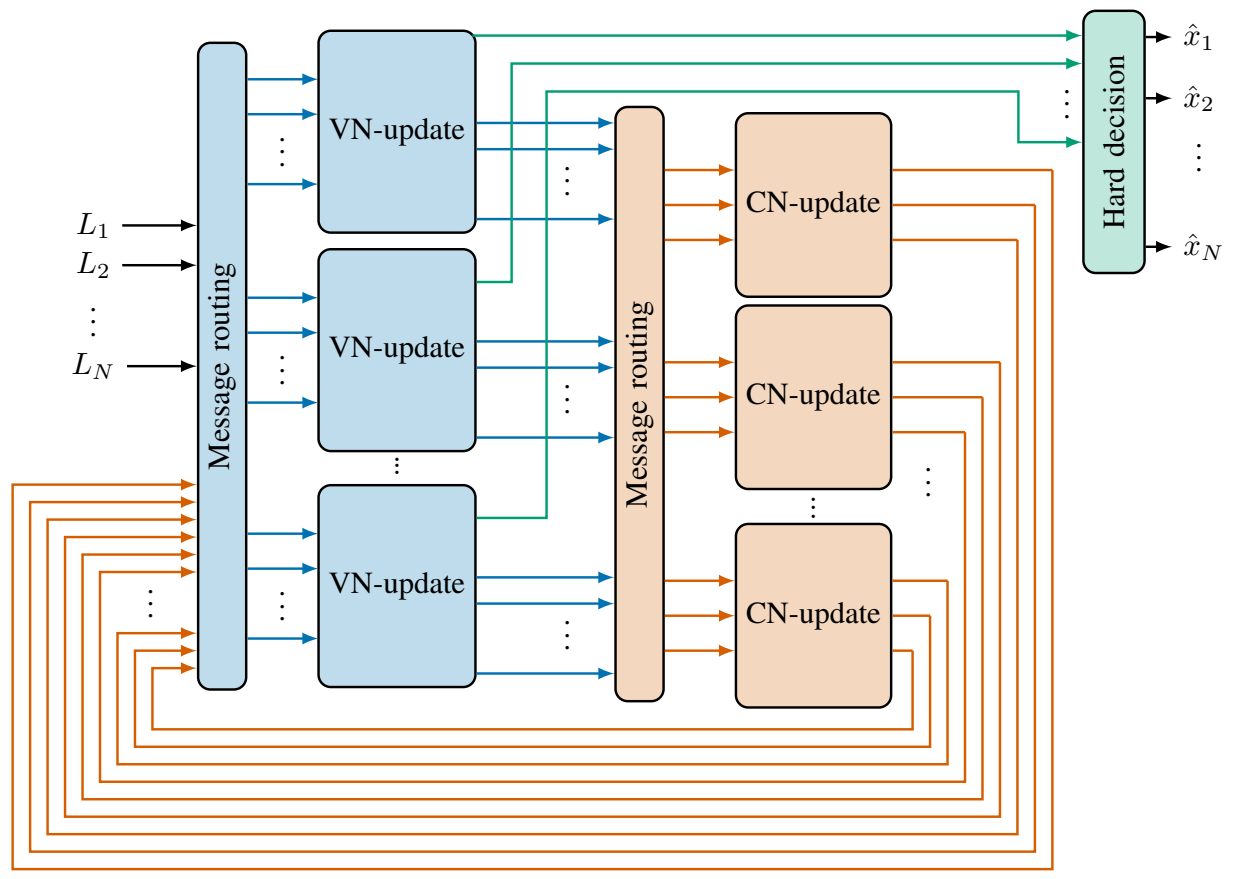
\begin{figure}
    \centering
    \resizebox{0.45\textwidth}{!}{\begin{tikzpicture}[>=latex, thick]
\def\ydist{0.5cm}
        \def\xdist{.8cm}
    
        \node [] (L1) {$L_1$};
        \node[below = -0.1cm of L1] (L2) {$L_2$};
        \node [below of =  L1] (h) {$\vdots$};
        \node[below = 1.2*\ydist of L2] (LN) {$L_N$};
    
        \node[right=1*\xdist of LN, draw, rectangle, rounded corners, align=center,minimum height = 7.4cm, fill=OIblue!25] (distVN) {\rotatebox{90}{Message routing}};  
        \draw[->] (L1) to (\tikztostart -| distVN.west);
        \draw[->] (L2) to (\tikztostart -| distVN.west);
        \draw[->] (LN) to (\tikztostart -| distVN.west);

        \def\h{0.7cm}
        \node [above right = -2.2cm and \xdist of distVN, draw, rectangle, rounded corners, minimum height = 3.3*\h, fill=OIblue!25] (VNU1) {VN-update};
        \node [above right = -4.7cm and \xdist of distVN, draw, rectangle, rounded corners, minimum height = 3.3*\h, fill=OIblue!25] (VNU2) {VN-update};
        \node [above right = -7.4cm and \xdist of distVN, draw, rectangle, rounded corners, minimum height = 3.3*\h, fill=OIblue!25] (VNUM) {VN-update};

        \def\yoff{0.4cm}
        \draw[<-, color=OIblue] (VNU1)+(-0.9cm,1.5*\yoff) to (\tikztostart -|  distVN.east);
        \draw[<-, color=OIblue] (VNU1)+(-0.9cm,0.5*\yoff) to (\tikztostart -|  distVN.east);
        \node [below right=-3.6*\yoff and -2.4cm of VNU1] {$\vdots$};
        \draw[<-, color=OIblue] (VNU1)+(-0.9cm,-1.5*\yoff) to (\tikztostart -|  distVN.east);

        \draw[<-, color=OIblue] (VNU2)+(-0.9cm,1.5*\yoff) to (\tikztostart -|  distVN.east);
        \draw[<-, color=OIblue] (VNU2)+(-0.9cm,0.5*\yoff) to (\tikztostart -|  distVN.east);
        \node [below right=-3.6*\yoff and -2.4cm of VNU2] {$\vdots$};
        \draw[<-, color=OIblue] (VNU2)+(-0.9cm,-1.5*\yoff) to (\tikztostart -|  distVN.east);

        \draw[<-, color=OIblue] (VNUM)+(-0.9cm,1.5*\yoff) to (\tikztostart -|  distVN.east);
        \draw[<-, color=OIblue] (VNUM)+(-0.9cm,0.5*\yoff) to (\tikztostart -|  distVN.east);
        \node [below right=-3.6*\yoff and -2.4cm of VNUM] {$\vdots$};
        \draw[<-, color=OIblue] (VNUM)+(-0.9cm,-1.5*\yoff) to (\tikztostart -|  distVN.east);
    
        \def\x{2.1cm}
        \node[above right=-2.5cm and (0.75*\x) of VNUM, draw, rectangle, rounded corners, minimum height = 6.8cm, fill=OIred!25] (dist) {\rotatebox{90}{Message routing}};

        \draw[->, color=OIblue] (VNU1.east)++(0cm, 0.1cm) to (\tikztostart -|  dist.west) node[above left=-1mm and 10mm] {};         
        \draw[->, color=OIblue] (VNU1.east)++(0cm,-0.2cm) to (\tikztostart -|  dist.west) node[above left=-1mm and 8mm](h1) {};
        \node [below left =-2.4mm and -0.7cm of h1] {$\vdots$};
        \draw[->, color=OIblue] (VNU1.east)++(0cm,-1.0cm) to (\tikztostart -|  dist.west) node[above left=-1mm and 2mm](h2) {};
  
        \draw[->, color=OIblue] (VNU2.east)++(0cm, 0.1cm) to (\tikztostart -|  dist.west) node[above left=-1mm and 10mm] {};         
        \draw[->, color=OIblue] (VNU2.east)++(0cm,-0.2cm) to (\tikztostart -|  dist.west) node[above left=-1mm and 8mm](h1) {};
        \node [below left =-2.4mm and -0.7cm of h1] {$\vdots$};
        \draw[->, color=OIblue] (VNU2.east)++(0cm,-1.0cm) to (\tikztostart -|  dist.west) node[above left=-1mm and 2mm](h2) {};

        \draw[->, color=OIblue] (VNUM.east)++(0cm, 0.1cm) to (\tikztostart -|  dist.west) node[above left=-1mm and 10mm] {};         
        \draw[->, color=OIblue] (VNUM.east)++(0cm,-0.2cm) to (\tikztostart -|  dist.west) node[above left=-1mm and 8mm](h1) {};
        \node [below left =-2.4mm and -0.7cm of h1] {$\vdots$};
        \draw[->, color=OIblue] (VNUM.east)++(0cm,-1.0cm) to (\tikztostart -|  dist.west) node[above left=-1mm and 2mm](h2) {};

        \node[below=-1.25mm of VNU2] () {$\cdot$};
        \node[below=-0.50mm of VNU2] () {$\cdot$};
        \node[below= 0.25mm of VNU2] () {$\cdot$};
        
        \def\h{0.7cm}
        \node [above right = -2.2cm and \xdist of dist, draw, rectangle, rounded corners, minimum height = 3.0*\h, fill=OIred!25] (CNU1) {CN-update};
        \node [above right = -4.4cm and \xdist of dist, draw, rectangle, rounded corners, minimum height = 3.0*\h, fill=OIred!25] (CNU2) {CN-update};
        \node [above right = -6.9cm and \xdist of dist, draw, rectangle, rounded corners, minimum height = 3.0*\h, fill=OIred!25] (CNUM) {CN-update};
        \node [below=-1mm of CNU2] (h2) {$\cdot$};
        \node [below= 0mm of CNU2] (h2) {$\cdot$};
        \node [below= 1mm of CNU2] (h2) {$\cdot$};

        \draw[<-, color=OIred] (CNU1)+(-0.9cm,\yoff) to (\tikztostart -| dist.east);
        \draw[<-, color=OIred] (CNU1)+(-0.9cm,0.0cm) to (\tikztostart -| dist.east);
        \draw[<-, color=OIred] (CNU1)+(-0.9cm,-\yoff) to (\tikztostart -| dist.east);

        \draw[<-, color=OIred] (CNU2)+(-0.9cm,\yoff) to (\tikztostart -| dist.east);
        \draw[<-, color=OIred] (CNU2)+(-0.9cm,0.0cm) to (\tikztostart -| dist.east);
        \draw[<-, color=OIred] (CNU2)+(-0.9cm,-\yoff) to (\tikztostart -| dist.east);
        
        \draw[<-, color=OIred] (CNUM)+(-0.9cm,\yoff) to (\tikztostart -| dist.east);
        \draw[<-, color=OIred] (CNUM)+(-0.9cm,0.0cm) to (\tikztostart -| dist.east);
        \draw[<-, color=OIred] (CNUM)+(-0.9cm,-\yoff) to (\tikztostart -| dist.east);

        \def\x{2.1cm}
        \node [above right=-2.8cm and 3.3*\x of VNU1, draw, rectangle, rounded corners, minimum height=3.0cm,minimum width =0.7cm, fill=OIgreen!25] (dec) {\rotatebox{90}{Hard decision}};

        \node [above right=-5.2mm and 3mm of VNU2] (n1) {};
        \node [above right=-5.2mm and 6mm of VNUM] (n2) {};

        \def\xoff{0.2cm}
        \draw[->, color=OIgreen] (VNU1.east)++(-0.05cm, 1.1cm) to (\tikztostart -| dec.west);  
        \draw[->, color=OIgreen] (VNU2.east)++(0cm, 7.8mm) -- +(2*\xoff,0cm) -- +(2*\xoff,6.25*\yoff) to (\tikztostart -| dec.west);
        \draw[->, color=OIgreen] (VNUM.east)++(0cm, 7.8mm) -- +(4*\xoff,0cm) -- +(4*\xoff,12.2*\yoff) -- +(6.2cm, 12.2*\yoff) -- +(6.2cm, 10.75*\yoff) to (\tikztostart -| dec.west);
        \node [below left = -2.45cm and 0.0cm of dec] {$\vdots$};

        \node[right=\x-0.6cm of CNU1] (m1) {};
        \node[right=\x-1.2cm of CNU2] (m2) {};
        \node[right=\x-1.8cm of CNUM] (m4) {};

        \draw[color=OIred] (CNU1) -- (m1.base) node[pos=0.4, above] () {}; 
        \draw[color=OIred] (m1.base)+(\xoff,\yoff) to (\tikztostart -| CNU1.east);
        \draw[color=OIred] (m1.base)+(-\xoff,-\yoff) to (\tikztostart -| CNU1.east);
        \draw[color=OIred] (CNU2) -- (m2.base) node[pos=0.4, above] (hc) {};
        \draw[color=OIred] (m2.base)+(\xoff,\yoff) to (\tikztostart -| CNU2.east);
        \draw[color=OIred] (m2.base)+(-\xoff,-\yoff) to (\tikztostart -| CNU2.east);
        \draw[color=OIred] (CNUM) -- (m4.base) node[pos=0.4, above] () {};
        \draw[color=OIred] (m4.base)+(\xoff,\yoff) to (\tikztostart -| CNUM.east);
        \draw[color=OIred] (m4.base)+(-\xoff,-\yoff) to (\tikztostart -| CNUM.east);
        \node[below of=hc] () {$\vdots$};
        
        \draw[->, color=OIred] (m1.base) -- +(0cm,-7.4cm) -- +(-11.5cm,-7.4cm) -- +(-11.5cm,-3.4cm) to (\tikztostart -| distVN.west);
        \draw[->, color=OIred] (m1.base)+(-\xoff,-\yoff) -- +(-\xoff,-7.4cm+\yoff/2) -- +(-11.5cm+\xoff,-7.4cm+\yoff/2) -- +(-11.5cm+\xoff, -3.4cm-\yoff/2) to (\tikztostart -| distVN.west);
        \draw[->, color=OIred] (m1.base)+(+\xoff,+\yoff) -- +(+\xoff,-7.4cm-\yoff/2) -- +(-11.5cm-\xoff,-7.4cm-\yoff/2) -- +(-11.5cm-\xoff, -3.4cm+\yoff/2) to (\tikztostart -| distVN.west);
        
        \draw[->, color=OIred] (m2.base) -- +(0cm,-4.6cm) -- +(-10.3cm,-4.6cm) -- +(-10.3cm,-1.8cm) to (\tikztostart -| distVN.west);
        \draw[->, color=OIred] (m2.base)+(-\xoff,-\yoff) -- +(-\xoff,-4.6cm+\yoff/2) -- +(-10.3cm+\xoff,-4.6cm+\yoff/2) -- +(-10.3cm+\xoff, -1.8cm-\yoff/2) to (\tikztostart -| distVN.west);
        \draw[->, color=OIred] (m2.base)+(+\xoff,+\yoff) -- +(+\xoff,-4.6cm-\yoff/2) -- +(-10.3cm-\xoff,-4.6cm-\yoff/2) -- +(-10.3cm-\xoff, -1.8cm+\yoff/2) to (\tikztostart -| distVN.west);
        
        \draw[->, color=OIred] (m4.base) -- +(0cm,-1.5cm) -- +(-9.1cm,-1.5cm) -- +(-9.1cm, -0.4cm) to (\tikztostart -| distVN.west);
        \draw[->, color=OIred] (m4.base)+(-\xoff,-\yoff) -- +(-\xoff,-1.5cm+\yoff/2) -- +(-9.1cm+\xoff,-1.5cm+\yoff/2) -- +(-9.1cm+\xoff, -0.4cm-\yoff/2) to (\tikztostart -| distVN.west);
        \draw[->, color=OIred] (m4.base)+(+\xoff,+\yoff) -- +(+\xoff,-1.5cm-\yoff/2) -- +(-9.1cm-\xoff,-1.5cm-\yoff/2) -- +(-9.1cm-\xoff, -0.4cm+\yoff/2) to (\tikztostart -| distVN.west);

        \node [below right = 3.55cm and 0.15cm of L1] {$\vdots$};

        \draw[->] (dec.east)++(0cm,1.2cm) -- +(0.3cm,0cm) node [right] (h4) {$\hat{x}_1$};
        \draw[->] (dec.east)++(0cm,0.5cm) -- +(0.3cm,0cm) node [right] {$\hat{x}_2$};
        \draw[->] (dec.east)++(0cm,-1.2cm) -- +(0.3cm,0cm) node [right] {$\hat{x}_N$};
        \node[below = 0.6cm of h4] () {$\vdots$};
\end{tikzpicture}}
    \caption{
        Fully node-parallel \gls{min-RCQ} decoder architecture.
    }
    \label{fig:arch:dec}
    \vspace*{-4mm}
\end{figure}

We implemented a fully node-parallel \gls{min-RCQ} decoder architecture that processes the iterations sequentially using the \glspl{LUT} design method following \cref{sec:Algorithm}. 
A sketch of the architecture is illustrated in \cref{fig:arch:dec}.
As shown in the center of \cref{fig:arch:dec}, $N$ \glspl{VN} and $M$ \glspl{CN} are instantiated which are used across all decoding iterations.
Two routing networks instantiate the Tanner graph edges to forward the \glspl{VN}-to-\glspl{CN} and \glspl{CN}-to-\glspl{VN} messages.
At the \gls{VN} outputs, a hard-decision unit applies a sign decision to the refined estimates $\tilde{L}_i$ (see \cref{eq:Estimate}) to obtain the decoded codeword bit $\hat{x}_i$.

The optimized quantization \glspl{LUT} are used at the Tanner graph edges connecting the \glspl{VN} to the \glspl{CN} of the respective degree.
The reconstruction \glspl{LUT} are used at the inputs of the \glspl{VN} connected to the \glspl{CN} of the respective degree.
Therefore, the \gls{CN} processing units in \cref{fig:arch:dec} execute \cref{eq:nms_cn2vn} with $\gamma=1$ on low-bit-width indices instead of high-precision \glspl{LLR}.
The \gls{VN} processing units perform the higher-precision update of~\cref{eq:rcq_vn2cn} (see \cref{fig:back:vnu_rcq}) and quantize outgoing messages to $q_\mathsf{l}$ bits.

Since the same \glspl{LUT} are reused across all iterations, no iteration-dependent multiplexers or control logic are required in the routing networks or processing units of \cref{fig:arch:dec}.
Thus, deriving \glspl{LUT} using \cref{alg:Algo1} enables the efficient implementation of non-unrolled, fully node-parallel \gls{min-RCQ} decoders.

\section{Results}
\label{sec:ds}
We design a \gls{min-RCQ} decoder using the proposed Bayesian optimization strategy for the $(1000,\,500,\,64)$ \gls{BG2} \gls{5G} \gls{LDPC} code~\cite{RK18}.
We compare both the error correction performance over a \gls{BI-AWGN} channel and the implementation complexity of this \gls{min-RCQ} decoder to a state-of-the-art \gls{NMS} decoder.

\subsection{Parameter Space for Integer to Integer Mapping}
\label{sec:ds:space}
For the $(1000,\,500,\,64)$ \gls{BG2} \gls{5G} \gls{LDPC} code, we set 
the channel quantizer $Q_{\Delta x}$ precision to $6$\,\unit{\bit}.
The precisions $q_\mathsf{h}$ and $q_\mathsf{l}$ are set to $6$\,\unit{\bit} and $4$\,\unit{\bit}, respectively. 
\cref{fig:space} shows the boundaries of the parameter subspace mentioned in \cref{sec:Algorithm:LUT_Design}.
The subspace between $R_{\textrm{s,LB}}$ and $R_{\textrm{s,UB}}$ is reduced by $42$\,\unit{\percent} compared to the overall search space between $R_{\textrm{LB}}$ and $R_{\textrm{UB}}$. 

The simulation of the transmission of codewords over a \gls{BI-AWGN} channel used a fixed 
${\frac{E_\mathsf{b}}{N_0}=2.5}$\,\unit{\dB}, and the number of transmitted codewords was limited to $C_\mathsf{w}=\num{1000000}$. The simulation was terminated when more than $50$ frame errors occurred.
The parameters contain the channel quantizer step size $\Delta x$ and the symmetric \glspl{LUT} of $Q_{i\rightarrow j}^{[\mathsf{v}]}$ and $R_{i\rightarrow j}^{[\mathsf{v}]}$ for~${d_\mathsf{c} = \{4, 5, 6, 7, 8\}}$.
This results in $36$ parameters in total. 
The quantizer step size $\Delta x$ was restricted to be in $\{0.25, 0.5, 1.0\}$, whereas the values of the \glspl{LUT} $Q_{i\rightarrow j}^{[\mathsf{v}]}$ and $R_{i\rightarrow j}^{[\mathsf{v}]}$ were bounded to the subspace shown in \cref{fig:space} under the monotonicity constraint.

For comparison, an \gls{NMS} decoder with a quantization of $6$\,\unit{\bit} was used.
This quantization was used for all \glspl{CN}, \glspl{VN}, and input \glspl{LLR}.
The channel quantizer step size was set to $\Delta x = 0.5$, and the extrinsic scaling factor $\gamma$ was set to $0.75$.

\subsection{Error Correction Performance}
\label{sec:ds:ecp}
\begin{figure}[t]
    \centering
    \resizebox{0.75\linewidth}{!}{\input{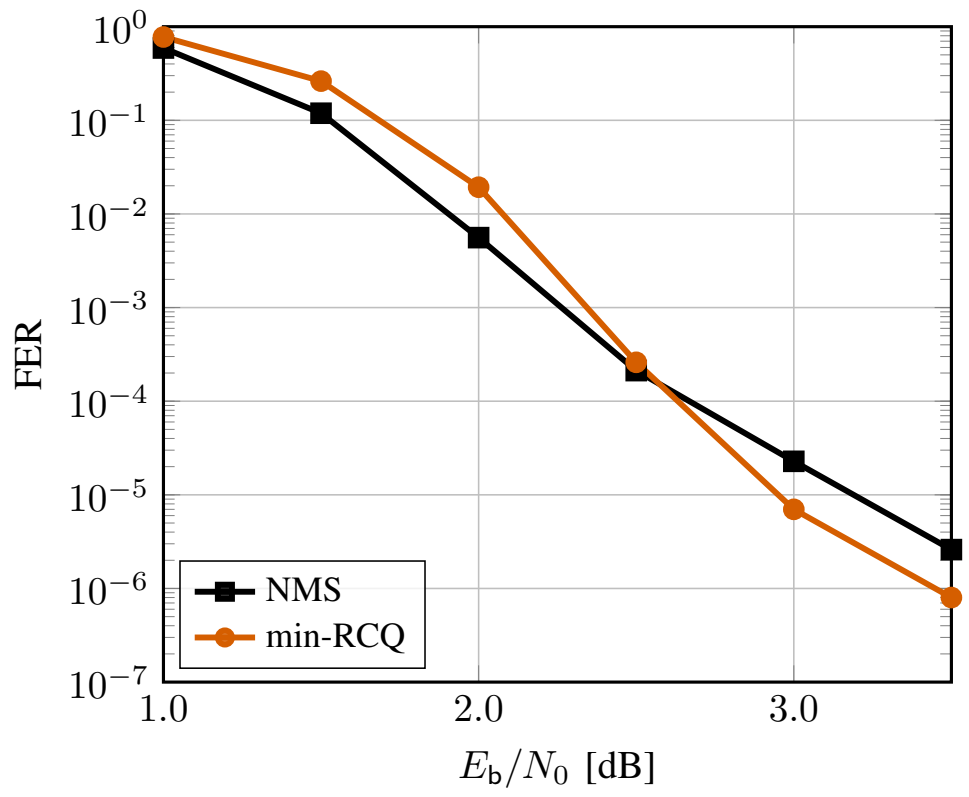}}
    \caption{\gls{FER} performance for the $(1000,\,500,\,64)$ BG2 5G \gls{LDPC} code. The \gls{FER} curves show the \gls{NMS} decoder and \gls{min-RCQ} decoder optimized with Bayesian optimization.}
    \label{fig:res_5g}
    \vspace*{-4mm}
\end{figure}

The error correction performance of the two decoders is shown in \cref{fig:res_5g}.
Both decoders perform $I = 32$ decoding iterations.
The \gls{min-RCQ} decoder provides an \gls{FER} performance close to the \gls{NMS} decoder. 
Below an $\frac{E_\mathsf{b}}{N_0}$ of $2.25$\,\unit{\dB}, the \gls{min-RCQ} decoder performs slightly worse than the \gls{NMS} decoder.
Above an $\frac{E_\mathsf{b}}{N_0}$ of $2.25$\,\unit{\dB}, the \gls{min-RCQ} decoder performs better than the quantized \gls{NMS} decoder.

This is expected, as the \gls{min-RCQ} decoder can outperform the quantized \gls{NMS} decoder~\cite{WWSB2020}. 
However, our approach uses the same \glspl{LUT} for all iterations, in contrast to \cite{WWSB2020}, which employs new \glspl{LUT} in each iteration.
Hence, potentially resulting in a degradation in error correction performance.

\subsection{Implementation using a $12$\,\unit{\nano\meter} Technology Node}
\label{sec:ds:imp}
In this section, we present virtual silicon implementation results in a $12$\,\unit{\nano\meter} \gls{FinFET} technology node for the \gls{min-RCQ} and \gls{NMS} decoders (see \cref{sec:ds:ecp}).
We use the following implementation metrics for comparison:
\begin{itemize}
    \item Clock frequency: $f$ [\unit{\mega\hertz}]
    \item Throughput: $T_\mathsf{c} = f \cdot N / I$ [\unit{\giga\bit\per\second}]
    \item Latency: $2 \cdot I / f$ [\unit{\nano\second}] (all designs have two pipeline stages, i.e., one register after each node update unit)
    \item Silicon area: $A$ [\unit{\milli\meter\squared}]
    \item Power consumption: $P$ [\unit{\watt}]
    \item Area efficiency: $\mu_\mathsf{a} = T_\mathsf{c} / A$ [\unit{\giga\bit\per\second\per\milli\meter\squared}]
    \item Energy efficiency:  $\mu_\mathsf{e} = P / T_\mathsf{c}$ [\unit{\pico\joule\per\bit}]
    \item Power density:  $\mu_\mathsf{e} \cdot \mu_\mathsf{a}$ [\unit{\watt\per\milli\meter\squared}]
\end{itemize}
Synthesis and \gls{PR} were performed with \textit{Synopsys Design Compiler} and \textit{Synopsys IC Compiler}, respectively.
A super-low $V_{t}$ library under worst-case \gls{PVT} conditions ($0.72$\,\unit{\volt}, $125$\,\unit{\celsius}) for timing and nominal case \gls{PVT} ($0.8$\,\unit{\volt}, $25$\,\unit{\celsius}) for power was used.
Power was extracted from post \gls{PR} netlists with back-annotated wiring. 
For a fair comparison, the cell utilization of both decoders was fixed to $70$\,\unit{\percent} with a target frequency of $600$\,\unit{\mega\hertz}.
Detailed implementation results are presented in \cref{tab:results}.
It can be seen that the \gls{min-RCQ} yields improved area efficiency and energy efficiency compared to the \gls{NMS} decoder.
Energy efficiency improves by $9.66$\,\unit{\pico\joule\per\bit} ($-38.6$\,\unit{\percent}), and area efficiency improves by $18.0$\,\unit{\giga\bit\per\second\per\milli\meter\squared} ($+24.0$\,\unit{\percent}).
These savings are within the expected range as they align with the reduction of wires connecting \glspl{VN} and \glspl{CN}.
The number of wires is reduced by $33.3$\,\unit{\percent} for the \gls{min-RCQ} with exchange messages quantized with $4$\,\unit{\bit} compared to the \gls{NMS} with exchange messages quantized with $6$\,\unit{\bit}.

\begin{table}[t]
    \centering
    \caption{
        Post \gls{PR} results using a $12$\,\unit{\nano\meter} \gls{FinFET} technology node.
    }
    \label{tab:results}
    \begin{tabular}{lcc}
    \hline
        Decoder                                                                 & NMS   & min-RCQ   \\
    \hline                                                                       
        Frequency [\unit{\mega\hertz}]                                          & $586$   & $600$       \\
        Throughput [\unit{\giga\bit\per\second}]                                & $21.1$  & $21.6$      \\
        Latency [\unit{\nano\second}]                                           & $109.2$ & $106.7$     \\
        Area [\unit{\milli\meter\squared}]                                      & $0.281$ & $0.232$     \\
        Power Consumption [\unit{\watt}]                                        & $0.528$ & $0.332$     \\
        Area Efficiency [\unit{\giga\bit\per\second\per\milli\meter\squared}]   & $75.1$  & $93.1$      \\
        Energy Efficiency [\unit{\pico\joule\per\bit}]                          & $25.02$ & $15.36$     \\
        Power Density [\unit{\watt\per\milli\meter\squared}]                    & $1.88$  & $1.43$      \\
    \hline
\end{tabular}
    \vspace*{-\PicSpace}
\end{table}

\subsection{Discussion}
\label{sec:ds:disc}
The results of the design study show improved \gls{FER} at high \gls{SNR} compared to an \gls{NMS} decoder with the same channel quantizer and internal \gls{VN} quantization. 
The \gls{min-RCQ} decoder has lower implementation complexity due to reduced quantization of the exchanged messages.
To enable the use of Bayesian optimization, the parameter space has been significantly reduced. 
However, this limits the degrees of freedom and, hence, may degrade the \gls{FER} performance. 
This quantification is beyond the scope of this paper.

\section{Conclusion}

We have proposed a method to optimize the parameters of a \gls{min-RCQ} decoder using Bayesian optimization.
To apply Bayesian optimization, we coupled the parameters of several update nodes and reduced the search space. 
Furthermore, we took no assumptions about the absence of cycles.
Neither do puncturing, shortening, or irregular codes impair the design method.
The design method can also be applied directly to hardware, since the optimizer only needs access to the parameters to be optimized, the codeword, and the output of the decoder.
The obtained parameter sets do not change over the iterations and, hence, can be used efficiently for fully node-parallel decoders.
To conclude, the proposed design method is simple, robust, and yields proper performance.

\newpage
\bibliographystyle{IEEEtran}
\bibliography{IEEEabrv,references}

@article{G62,
  author    = {Robert G. Gallager},
  title     = {{Low-Density Parity-Check Codes}},
  journal   = {IEEE Trans. Inf. Theory},
  volume    = {8},
  number    = {1},
  pages     = {21--28},
  month     = jan,
  year      = {1962}
}

@ARTICLE{RK18,
  author={Richardson, Tom and Kudekar, Shrinivas},
  journal={IEEE Commun. Mag.}, 
  title={{Design of Low-Density Parity Check Codes for 5G New Radio}}, 
  year={2018},
  volume={56},
  number={3},
  pages={28-34},
  doi={10.1109/MCOM.2018.1700839}
}

@techreport{3GPP26,
  author    = {{3GPP}},
  title     = {{NR; Multiplexing and Channel Coding}},
  institution = {3GPP},
  number    = {TS 38.212 V19.3.0},
  month     = mar,
  year      = {2026}
}

@standard{IEEE21,
  title     = {{Wireless LAN Medium Access Control (MAC) and Physical Layer (PHY) Specifications – Amendment 1: Enhancements for High-Efficiency WLAN}},
  organization = {IEEE},
  number    = {IEEE Std 802.11-2020},
  year      = {2021}
}

@book{RU2008,
  author    = {Thomas J. Richardson and R{\"u}diger L. Urbanke},
  title     = {{Modern Coding Theory}},
  publisher = {Cambridge University Press},
  year      = {2008}
}

@article{RS01,
  author    = {Thomas J. Richardson and Mohammad A. Shokrollahi and R{\"u}diger L. Urbanke},
  title     = {{Design of Capacity-Approaching Irregular Low-Density Parity-Check Codes}},
  journal   = {IEEE Trans. Inf. Theory},
  volume    = {47},
  number    = {2},
  pages     = {619--637},
  month     = feb,
  year      = {2001}
}

@article{FMI99,
  author    = {Marc P. C. Fossorier and M. Mihaljevic and Hideki Imai},
  title     = {{Reduced Complexity Iterative Decoding of Low-Density Parity-Check Codes Based on Belief Propagation}},
  journal   = {IEEE Trans. Commun.},
  volume    = {47},
  number    = {5},
  pages     = {673--680},
  month     = may,
  year      = {1999}
}

@ARTICLE{CDEFH2005,
  author={Chen, J. and Dholakia, A. and Eleftheriou, E. and Fossorier, M.P.C. and Hu, X.Y.},
  journal={IEEE Trans. Commun.}, 
  title={{Reduced-complexity decoding of LDPC codes}}, 
  year={2005},
  volume={53},
  number={8},
  pages={1288-1299}}

@article{PDDV2013,
  author    = {S. K. Planjery and D. Declercq and L. Danjean and B. Vasic},
  title     = {{Finite Alphabet Iterative Decoders—Part I: Decoding Beyond Belief Propagation on the Binary Symmetric Channel}},
  journal   = {IEEE Trans. Commun.},
  volume    = {61},
  number    = {10},
  pages     = {4033--4045},
  month     = oct,
  year      = {2013}
}

@article{GBMMMTB2018,
  author    = {R. Ghanaatian and A. Balatsoukas-Stimming and T. C. M{\"u}ller and M. Meidlinger and G. Matz and A. Teman and A. Burg},
  title     = {{A 588-Gb/s LDPC Decoder Based on Finite-Alphabet Message Passing}},
  journal   = {IEEE Trans. Very Large Scale Integr. (VLSI) Syst.},
  volume    = {26},
  number    = {2},
  pages     = {329--340},
  year      = {2018}
}

@inproceedings{WWSB2020,
  author    = {L. Wang and R. D. Wesel and M. Stark and G. Bauch},
  title     = {{A Reconstruction-Computation-Quantization ({RCQ}) Approach to Node Operations in LDPC Decoding}},
  booktitle = {Proc. IEEE Glob. Commun. Conf. (GLOBECOM)},
  pages     = {1--6},
  year      = {2020}
}

@article{SWBW2020,
  author    = {M. Stark and L. Wang and G. Bauch and R. D. Wesel},
  title     = {{Decoding Rate-Compatible 5G-LDPC Codes with Coarse Quantization Using the Information Bottleneck Method}},
  journal   = {IEEE Open J. Commun. Soc.},
  volume    = {1},
  pages     = {646--660},
  year      = {2020}
}

@inproceedings{M1974,
  author    = {J. Mockus},
  title     = {{On Bayesian Methods for Seeking the Extremum}},
  booktitle = {Optimization Techniques IFIP Tech. Conf.},
  year      = {1974}
}

@article{LU2018,
  author    = {J. Lewandowsky and G. Bauch},
  title     = {{Information-Optimum LDPC Decoders Based on the Information Bottleneck Method}},
  journal   = {IEEE Access},
  volume    = {6},
  pages     = {4054--4071},
  year      = {2018}
}

@article{JSW1998,
  author    = {D. R. Jones and M. Schonlau and W. J. Welch},
  title     = {{Efficient Global Optimization of Expensive Black-Box Functions}},
  journal   = {J. Global Optim.},
  volume    = {13},
  pages     = {455--492},
  year      = {1998}
}

@misc{scikit,
  author    = {T. Head and M. Kumar and H. Nahrstaedt and G. Louppe and I. Shcherbatyi},
  title     = {{scikit-optimize}},
  howpublished = {\url{https://github.com/scikit-optimize}},
  doi = {10.5281/zenodo.5565057},
  year      = {2021}
}

@misc{github-code,
  author    = {A. von Bank},
  title     = {{Bayesian Optimization for Finite Alphabet Message Passing Decoders}},
  howpublished = {\url{https://github.com/kit-cel/bayesian_minRCQ_opt}},
  year      = {2026}
}

@article{TG20,
  author = {T. Monsees and O. Griebel and M. Herrmann and D. Wübben and A. Dekorsy and N. Wehn},
  title = {{Minimum-Integer Computation Finite Alphabet Message Passing Decoder: From Theory to Decoder Implementations towards 1 Tb/s}},
  journal = {Entropy},
  volume = {24},
  year = {2022},
  number = {10},
  ARTICLE-NUMBER = {1452},
  ISSN = {1099-4300}
}

@article{mackay,
  author    = {D.J.C. MacKay and R.M. Neal},
  title     = {{Near Shannon Limit Performance of Low Density Parity Check Codes}},
  journal   = {Electron. Lett.},
  volume    = {32},
  number    = {18},
  pages     = {1645-1646},
  month     = aug,
  year      = {1996}
}

@book{RW2006,
  author    = {C. E. Rasmussen and C. K. I. Williams},
  title     = {{Gaussian Processes for Machine Learning}},
  publisher = {Massachusetts Institute of Technology Press},
  year      = {2006}
}

@inproceedings{GJKHW2024,
  author={Griebel, Oliver and Johannsen, Lucas and Kestel, Claus and Herrmann, Matthias and Wehn, Norbert},
  booktitle={Proc. Asilomar Conference on Signals, Systems, and Computers}, 
  title={{Channel Coding Under Silicon Implementation Constraints}}, 
  year={2024},
  volume={},
  number={},
  pages={967-974},
  doi={10.1109/IEEECONF60004.2024.10942759}}

\end{document}